\documentclass[12pt,fleqn]{article}
\usepackage{amsmath,a4,latexsym,epsfig}
\newcommand{\intd}{\int \! d^4 x \;}
\newcommand{\intS}{\int \! d S \;}
\newcommand{\intSbar}{\int \! d \bar S \;}
\newcommand{\intV}{\int \! d V \;}
\newcommand{\Ga} {\Gamma}
\newcommand{\Gacl} {{\Gamma_{\rm cl}}}
\newcommand{\equa}[1]{\begin{align} #1 \end{align}}

\renewcommand{\L}{{\cal L}}
\newcommand{\lambdabar}{{\overline\lambda}}
\newcommand{\sigmabar}{{\overline\sigma}}
\newcommand{\psibar}{{\overline\psi}}

\newcommand{\phibar}{{\overline \varphi}}
\newcommand{\epsilonbar}{{\overline\epsilon}}
\newcommand{\thetabar}{{\overline\theta}}
\newcommand{\etabar}{{\overline\eta}}
\newcommand{\chibar}{{\overline\chi}}
\newcommand{\qbar}{{\overline q}}
\newcommand{\fbar}{{\overline f}}
\newcommand{\cbar}{{\overline c}}

\newcommand{\alphadot}{{\dot\alpha}}
\newcommand{\betadot}{{\dot\beta}}

\newcommand{\lambdaV}{{\tilde \lambda}}
\newcommand{\lambdaVbar}{{\overline {\tilde \lambda}}}
\newcommand{\DV}{{\tilde D}}

\def\dF#1{\frac{\delta{\cal F}}{\delta#1}}
\def\dfunc#1#2{\frac{\delta^{#1}}{\delta#2}}

\def\pslash#1{{\setbox0=\hbox{$#1$}
  \rlap{\ifdim\wd0>.7em\kern.22\wd0\else\kern.1\wd0\fi /}#1}}

\newcommand{\mn}{\mu\nu}

\begin{document}
\begin{titlepage}

\begin{flushright}
KA--TP--14--2001\\
BN--TH--01--04\\
{\tt hep-th/0105028}\\
\end{flushright}

\begin{center}
{\Large\bf{
        Non-renormalization theorems of   Supersymmetric QED \\[1ex]
        in the Wess--Zumino gauge}}
\\
\vspace{8ex}
{\large         E. Kraus$^a$} and
{\large        D. St{\"o}ckinger$^b$}
{\renewcommand{\thefootnote}{\fnsymbol{footnote}}
\footnote{E-mail addresses:\\
                kraus@th.physik.uni-bonn.de,\\
                ds@particle.physik.uni-karlsruhe.de.}} 
\\
\vspace{2ex}
{\small\em               $^a$ Physikalisches Institut,
              Universit{\"a}t Bonn,\\
              Nu{\ss}allee 12, D--53115 Bonn, Germany\\
}
\vspace{.5ex}
{\small\em $^b$ Institut f{\"u}r Theoretische Physik, 
              Universit{\"a}t Karlsruhe,\\
              D--76128 Karlsruhe, Germany\\}
\vspace{2ex}
\end{center}
\vfill
{\small
 {\bf Abstract}
 \newline\newline
The non-renormalization theorem of chiral vertices and the
generalized non-renorma\-lization theorem of the photon self energy are
derived in SQED on the basis of  algebraic renormalization. For this
purpose the gauge coupling  is extended to an external
superfield. This extension already provides detailed insight into the
divergence structure.
Moreover, using  the local supercoupling together with an
additional external vector multiplet that couples to the axial
current, the model becomes complete in the sense of
multiplicative renormalization, with two important implications.
First, a Slavnov--Taylor identity describing
supersymmetry, gauge symmetry, and axial symmetry including the axial
anomaly can be established to all orders. Second, from this
Slavnov--Taylor identity we can infer a 
Callan--Symanzik equation expressing all aspects of the
non-renormalization theorems. In particular,  the gauge
$\beta$-function appears explicitely in the  closed form.

\vfill
}

\end{titlepage}


\newpage
\section{Introduction}

Supersymmetric theories have always been famous for their
extraordinary  renormalization properties. Particular
non-renormalization theorems state the absence of divergences to the
superpotential \cite{FULA75, GSR79},
 and the $\beta$-functions of supersymmetric gauge
theories have been given in a closed form as a function of a one-loop
coefficient and the anomalous dimension of the matter fields
\cite{NSVZ83, SHVA86}. However,
non-renormalization theorems have been derived in general only as a
consequence of the Feynman rules in superspace. Outside of superspace,
the non-renormalization theorems have not been proven. And the
expressions for the gauge $\beta$-function can be derived
perturbatively 
 only 
 by constructing the supercurrent in the
manifestly supersymmetric gauge and in a strict sense they are only
available
in supersymmetric QED (SQED) \cite{CPS79, LPS87}. In this
construction,  the 
relation to the divergence structure of underlying diagrams is not
apparent.

Owing to the increasing number of phenomenological calculations carried
out in the Wess--Zumino gauge the situation is not very
satisfactory. From explicit calculations in the Wess--Zumino gauge it
was apparent that the non-renormalization theorems for the
superpotential do not hold in the form derived in superspace
\cite{WZ74_SQED}.
 However,
it was argued that their consequences for gauge independent quantities
should hold in the Wess--Zumino gauge as well. For example, the
$\beta$-functions of chiral couplings and masses in the Callan--Symanzik
equation should be restricted in the same way as in superspace, where
they are related to each other by a gauge independent field
renormalization.   Since a transition from the supersymmetric gauge to
the Wess--Zumino gauge cannot be  performed in perturbation 
theory, it is important to derive the results directly in the
Wess--Zumino gauge by using algebraic properties of the supersymmetric actions.

Such an algebraic derivation of the non-renormalization theorems
has been performed in a foregoing paper in the context of the
Wess--Zumino model \cite{FLKR00}. In the present paper we will apply the same
analysis to the supersymmetric extension of QED in the Wess--Zumino
gauge \cite{WZ74_SQED} and derive both the non-renormalization
theorems of chiral 
vertices and the closed form of the $\beta$-function.
 
The algebraic analysis is based on the following observations:
Supersymmetric Lagrangians can be written as the highest component of
a supermultiplet, and so they are related to  lower dimensional field
monomials by supersymmetry transformations. The implications of this
multiplet structure for Green functions can be worked out by extending
the coupling constant to an external superfield. Differentiation with
respect to the local supercoupling yields Green functions with one insertion of
the supermultiplet of the interaction Lagrangian. For constant coupling
these Green functions are the ones of the original model. But now they
are related to Green functions with lower-dimensional vertex
insertions by supersymmetry. In the context of the Wess-Zumino model it
was shown that 
these relations imply an improvement for the power-counting degree of
divergence for chiral Green functions.

In the algebraic approach non-renormalization theorems can be first
considered on the basis of invariant counterterms. Whenever invariant
counterterms are forbidden 
for reasons of symmetry, the corresponding Green functions are related
to non-local expressions. Hence, absence of counterterms can be viewed as a
manifestation of underlying non-renormalization theorems. These
non-renormalization theorems can be worked out by relating the
respective Green functions to the non-local expressions.

When we extend the
gauge coupling of SQED to an external superfield, it turns out that
 independent counterterms to chiral vertices and counterterms to the
photon self energy from two-loop order onwards are absent. 
However, in the Wess--Zumino gauge the non-linear supersymmetry permits
 individual field renormalizations for
all matter  fields. Therefore, we obtain the non-renormalization
theorem of chiral vertices similar to the one present in superspace
but modified by these gauge dependent field renormalizations.

By working out the non-renormalization theorems in explicit
expressions, we find that the
non-renormalization theorem of chiral vertices and the generalized
non-renormalization theorem of the photon self energy
 are of a different nature:  Chiral Green functions are
superficially convergent up to  gauge-dependent field redefinition;
in contrast, the photon self energy is related to linearly divergent Green
functions, which become meaningful only in the course of
renormalization. Their local divergent part, however, is
uniquely determined from  the non-local one by gauge invariance, and
it is for this reason that counterterms representing the independent
divergences cannot appear in higher orders.

In the past it could only be suggested from the closed
expression of the gauge $\beta$-function that there is an underlying
generalized non-renormalization theorem for the photon self energy.
Vice versa,  we have to prove in the present context, how the
generalized non-renormalization theorem of the photon self energy
gives rise to the closed form of the gauge $\beta$ function. For this
purpose we have to derive the Callan--Symanzik equation of SQED with
local gauge coupling.

The  Callan--Symanzik equation can only
be derived if all invariant counterterms can be understood as field
and coupling redefinitions 
--- which  is
not the case in the presence of local couplings. For this reason  we introduce
an axial vector multiplet whose vector component
couples to the  axial current and gives rise to an (anomalous) axial
Ward identity. Combining axial symmetry and local gauge coupling, the
construction of the theory is remarkable on both sides: For local
couplings, the model is multiplicatively renormalizable only when the
axial vector multiplet is introduced; for axial symmetry the
Adler--Bardeen anomaly can be absorbed  into  the Slavnov--Taylor 
operator by means
of the local coupling and the model can be constructed by algebraic
renormalization in the presence of the anomaly. Even the
non-renormalization of the Adler--Bardeen anomaly \cite{ADBA69} is a simple
consequence of the local coupling. 

By using the extended action with the 
gauged axial current and the local coupling the Callan--Symanzik equation
can be derived and describes consistently the scaling of the local
coupling and of the axial vector multiplet in presence of the
anomaly. The closed form of the gauge $\beta$-function is the result
of the algebraic construction of the Callan--Symanzik equation as a linear
combination of symmetric operators with respect to the anomalous
Slavnov--Taylor identity.

According to the general outline the paper is divided into two parts:
In the first part of the paper (section 2--5) we derive the
non-renormalization theorems of chiral vertices and the photon self energy:
In section 2  SQED is extended to
a supersymmetric theory with local coupling, in section 3 we outline
the renormalization of SQED with local couplings in the Wess--Zumino
gauge.  In section 4 we construct the invariant counterterms and find
the non-renormalization theorems implicitly as absence of
counterterms to chiral vertices and to the photon self-energy from
2-loop order onwards. In section 5 the analysis is continued to the
explicit construction of the corresponding non-local expressions.
 
The second part (section 6--9) is devoted to the derivation of the Callan-Symanzik
equation  and, in particular, of
the closed form of the $\beta$-function:
We introduce the axial vector multiplet  in section 6
and construct the Slavnov--Taylor identity in presence of the
Adler--Bardeen anomaly in section 7.  Finally, in section 8 we derive
the Callan--Symanzik equation, and we find the implications of
non-renormalization theorems as restrictions on the Callan--Symanzik
coefficients. In section 9 we derive an interesting  relation between
the axial-current Green functions and the photon self energy. This
relation explains the appearance of terms of higher orders to
the gauge $\beta$-function and can be identified as the analogue to the
Konishi anomaly \cite{CPS79, KO81} in the Wess--Zumino gauge. In
the appendices we give 
the notations and  conventions,  the BRS transformations and
the transformations of fields under discrete symmetries.


\section{SQED with local couplings }
\label{SecDefinition}

In the manifestly supersymmetric gauge it is obvious that the 
Lagrangian of an invariant action is the highest component of a
supermultiplet. As will be shown here, the gauge invariant parts
of  the Lagrangian
of    SQED in the Wess--Zumino gauge are  
the highest components of ordinary supermultiplets, too. This observation
is the basis for the derivation of the non-renormalization
theorems. Technically it is exploited by extending the gauge coupling to a space-time
dependent external field, the local coupling. In order to maintain
supersymmetry, the local gauge coupling has to be taken as 
the lowest component of a constrained real superfield.
The invariant action then includes as
additional terms the complete
chiral and antichiral multiplet of the gauge and supersymmetric
invariant kinetic Lagrangian of the photon multiplet, whose lowest
component is the photino mass term.

\subsection{The multiplet structure of the gauge invariant action}

 The classical action of the supersymmetric extension of QED (SQED)
 \cite{WZ74_SQED}   extends the gauge invariant action of ordinary QED
to a supersymmetric  action. In the Wess--Zumino gauge it
 contains the vector multiplet $(A^ \mu, \lambda^ \alpha, \lambdabar^
 \alphadot, D)$ and the left and right handed chiral 
 multiplets
$(\varphi_L, \psi_L^ \alpha, F_L)  $ and $(\varphi_R, \psi_R^ \alpha,
 F_R)$
and the respective complex conjugate antichiral
 multiplets.\footnote{For the purpose of the present section we 
keep the auxiliary fields $D$ and $F$ in the action, and we eliminate
them when we proceed to quantization in section 3.}
 The matter fields
are charged with the electric charge $Q_L= -1 $  and $Q_R = 1 $.

The invariant action can be decomposed into the invariant kinetic part
of the photon and photino, the matter part containing the interaction
of the matter fields with the photon multiplet and the supersymmetric
mass term of matter:
\begin{eqnarray}
\Ga_{\rm SQED} & = &\Ga_{\rm kin} + \Ga_{\rm matter} + \Ga_{\rm mass}
\nonumber \\
& = &  \intd \Bigl(\frac 12 (L_{\rm kin} + \bar L_{\rm kin}) + L_{\rm matter} +
                       m (  L_{\rm mass} + {\bar L}_{\rm mass}) \Bigr)
\label{Gasqed}
\end{eqnarray}
The corresponding Lagrangians 
are defined by the following  expressions:
\begin{eqnarray}
\label{Lkin}
L_{\rm kin} & = & - \frac 14 F^{\mn} F_{\mn} +  i \lambda^ \alpha
 \sigma^\mu_{\alpha\alphadot} \partial_\mu\lambdabar ^ {\alphadot} + \frac 1 8 
D^2
- \frac i 8 \epsilon^{\mu \nu \rho \sigma} F_{\mn} F_{\rho \sigma}
 \\
\label{Lmass}
L_{\rm mass} & = & F_L \varphi_R + F_R \varphi_L - \psi_L \psi_R 
\end{eqnarray}
and respective expressions for their complex conjugates.
$L_{\rm matter}$ can be split into a  left- and right-handed part:
\begin{equation}
\label{Lmatter1}
{L}_{\rm matter}  =  L_{\rm matter, L} +  L_{\rm matter, R}
\end{equation}
where ($A =  L,R$)
\begin{eqnarray}
L_{\rm matter, A}
& = &\frac 12
D^ \mu  \phibar_A D_\mu \varphi_A - \frac 1 4  \phibar_A D^ \mu D_\mu \varphi_A
-\frac 1 4  \varphi_A D^ \mu D_\mu  \phibar_A \nonumber \\
& & \; + \frac i 2  \psi_A^ \alpha \sigma_{\alpha \alphadot}  D_\mu \psibar_A
^ {\alphadot}
-  \frac i 2  D _\mu \psi_A^ \alpha \sigma^\mu 
          _{\alpha \alphadot}   \psibar_A^ {\alphadot}
+ {\overline F}_A  F_A \nonumber \\
& & \;+ i e Q_A  \sqrt{2}( \lambda \psi_A  \phibar_A  -  \lambdabar  \psibar_A
 \varphi_A)
+ \frac 1 2 e Q_A D \phibar_A \varphi _A
\label{Lmatter}
\end{eqnarray}
 The  field strength $F_{\mn}$ and the  gauge covariant derivative  
are given by
\begin{eqnarray}
\label{covD}
D_\mu \phi_A & = & (\partial_\mu+ieQ_A  A_\mu)\phi_A\ , \qquad 
D_\mu{\bar  \phi}_A  =  (\partial_\mu-ieQ_A A_\mu){\bar \phi_A} \ ,\nonumber \\
F_{\mu\nu}(A) & = & \partial_\mu A_\nu - \partial_\nu A_\mu\ .
\end{eqnarray}

In the Wess--Zumino gauge   
 the algebra of supersymmetry transformations
closes
on translations only up to an abelian gauge transformation  $\delta^
 {\rm gauge}_\omega$ :
\begin{equation}
\label{susyalg}
\{\delta_\alpha , \bar \delta_\alphadot\} = 2i \sigma^ \mu_{\alpha
\alphadot} \bigl(\partial_\mu +  \delta^ {\rm gauge}_{eA^ \mu}\bigr)
\end{equation}
When applied to 
gauge invariant expressions 
 the algebra closes on translations.
The Lagrangians above are gauge invariant and transform covariantly 
under supersymmetry transformations. Hence, the supersymmetry
algebra implies that 
they can be
written as a  supersymmetry variation of lower dimensional field monomials:
 In fact, $L_{\rm  kin} $ (\ref{Lkin}) and 
$L_{\rm mass}$  (\ref{Lmass}) are the highest
components of  chiral multiplets, and as such they are the
second  variations of their lowest
components:
\equa{\label{Lkinch}
{\L}_{\rm kin} (x,\theta) &  = - \frac 12 
\lambda^{ \alpha} \lambda _\alpha - \frac i2 \theta^
\alpha 
( \sigma^{\mn
\; \beta} _\alpha \lambda _\beta F_{\mn} +  \lambda _\alpha D) +
\theta^ 2 L_{\rm kin}(x) \\
\label{Lmassch}
{\L}_{\rm mass}(x,\theta) & = \varphi_L \varphi_R + \sqrt{2} \theta^ \alpha
(\psi_{L\; \alpha} \varphi_R + \psi_{R \; \alpha} \varphi_L) + 
\theta^ 2 L_{\rm mass}(x)
}
Their transformations  under supersymmetry can be given in a closed
form by using the superspace formulation (see (\ref{chiral} and
(\ref{susychiral}) in appendix A for definitions and  conventions):
\begin{equation}
\label{Lchsusy}
\delta_\alpha {\L} (x, \theta) = \frac {\partial}{\partial \theta^
\alpha} {\L} 
(x, \theta) \quad
\bar \delta_\alphadot {\L} (x, \theta) = -2i (\theta \sigma ) _{\alphadot}
\partial_{\mu}{\L} (x,\theta)
\end{equation}  
$\bar L_{\rm kin} $ and $\bar L_{\rm mass} $ are the highest components of the
 respective antichiral multiplets.
The matter part of the action is the highest component of a real
supermultiplet and can be written even as a fourth variation of the
2-dimensional field monomial $\varphi_A \phibar_A$ (see (\ref{Lmattervek})).

In the following section we show that
the chiral multiplet ${\cal L}_{\rm kin}$ and its complex conjugate
determine the supersymmetric extension of the local gauge coupling to
a real  superfield, which is composed of a chiral and antichiral
superfield.

\subsection{SQED with local gauge  coupling}

In a first step towards a supersymmetric action with local gauge
 coupling, we
 extend the   coupling constant $e$ in the matter Lagrangian (\ref{Lmatter})
  to an external field $e(x)$:
\begin{equation}
\label{Lmatterex}
e\to e(x) \quad 
\mbox{\rm and} \quad L_{\rm matter}(e) \to L_{\rm matter} (e(x))\ .
\end{equation} 
It is gauge invariant with local coupling $e(x)$ under modified gauge
transformations:
\equa{\label{gaugetr}
\delta^ {\rm gauge} A_{\mu} &= \frac 1 {e(x)} \partial_\mu \omega (x), \qquad  
 \delta^ {\rm gauge}\lambda  = \delta^ {\rm gauge} D  = 0, \\
\delta^ {\rm gauge} \phi_A & = -iQ_A \omega(x) \phi_A,\quad
\delta^ {\rm gauge} {\overline \phi}_A  = iQ_A \omega(x) {\overline \phi}_A,
\qquad \phi =
\varphi, \psi, F. \nonumber  
}
$\Ga_{\rm matter}(e(x))$ is even invariant under  supersymmetry
transformations, if the supersymmetry transformations are modified to
\equa{
\delta_\alpha( e A^ \mu) & = i \sigma_{\alpha \alphadot} ^ \mu e \lambdabar^ 
\alphadot \ ,&  &\bar \delta_{\alphadot}
(eA^ \mu)  = - i e \lambda^  
\alpha\sigma_{\alpha \alphadot} ^ \mu\ , \nonumber\\ 
\delta_\alpha (e\lambda^ \beta) & = 
\frac i 2 \sigma^{\mn \; \beta}_\alpha  F_{\mn}(eA)
+ \frac i 2 \delta_\alpha^ \beta e D , &  &\bar \delta_{\alphadot} (e\lambda 
_{\beta}) = 0\ ,\nonumber \\
\delta_\alpha (e\lambdabar_\alphadot) & = 0\ ,  &  &\bar \delta^ {\alphadot}
(e\lambdabar _{\dot \beta})=\frac i 2 {\bar \sigma}^ {\mn \; \alphadot }_
{\phantom{\mn; \alpha} \dot \beta}
  F^ {\mn} (eA)
-  \frac  i 2   \delta^ \alphadot_ {\dot\beta}e D , \nonumber \\
\delta_\alpha (e D) & = 
2 \sigma^ \mu_{\alpha \alphadot} \partial _\mu (e\lambdabar ^ 
\alphadot) \ , &  &\bar \delta_{\alphadot}(e D) = 
2 \partial _\mu (e\lambda ^ 
\alpha)\sigma^ \mu_{\alpha \alphadot}\ ,
\label{susyvector}
}
and ($A = L,R$)
\equa{
\delta_{\alpha} \varphi_A & = \sqrt{2} \psi_{A\;\alpha}\ , 
& \quad &  \bar\delta_{\alphadot} \varphi_A = 0\ , \nonumber\\
\delta_{\alpha} \psi_A^ {\beta} & = \sqrt{2} \delta_\alpha^ \beta F_A\ , 
& \quad &  \bar \delta_{\alphadot} \psi_{A\;\alpha} = \sqrt{2}i 
\sigma^ \mu_{\alpha\alphadot}D_\mu \varphi_A\ , \nonumber\\
\delta_{\alpha}  F_A & = 0 \ ,& \quad  &  \bar \delta_{\alphadot}F_A  = 
\sqrt{2}i D_\mu \psi_A^ \alpha \sigma^ \mu_{\alpha\alphadot} + 2 i e Q_A
 \lambdabar _ \alphadot  \varphi_A\ .
\label{susymatter}
}
Since the gauge transformation of the field $e(x) A^
\mu$ takes the same form as the gauge transformation with constant coupling,
   the
supersymmetry transformations in terms of the fields 
 $(e(x) A^ \mu, e(x)\lambda , e(x) \lambdabar,
e(x) D)$  take the same form as the usual supersymmetry transformations.
These transformations are in
agreement with the algebraic restrictions, since the
supersymmetry transformations commute with the gauge
transformations (\ref{gaugetr})  and
fulfill the   supersymmetry algebra as given in
(\ref{susyalg}). 

$\Ga_{\rm matter}$ depends on the coupling only via the product with a
photon, photino or D-field. For this reason it is invariant under the
supersymmetry transformations with local couplings no matter how the
transformation of $e(x)$ itself is defined.

However, the extension of  $\Ga_{\rm kin}$ to a gauge invariant
expression  with local gauge
coupling,
\begin{equation} 
\label{Lgaugeinvariant}   
\intd \Bigl( - \frac 1{4e^2}F^ {\mn}(eA) F_{\mn} (eA) + \frac i2 (
\lambda \sigma \partial \lambdabar - \partial \lambda \sigma
\lambdabar)+\frac 1 8 D^2 \Bigr),
\end{equation}
 depends on the coupling
$e(x)$ not only in the combination with the photon field and is not
supersymmetric without further modifications. This is particularly
transparent in the superspace formulation of (\ref{Lgaugeinvariant}),
\begin{equation}
\intS \frac{1}{2e(x)^2} \L_{\rm kin}
 + \intSbar\frac{1}{2e(x)^2} \overline{\L}_{\rm kin}\, .
\label{LgaugeinvariantS}
\end{equation}
Therefore, in a second step we extend the coupling $e(x)$ to a
supermultiplet of fields.


Indeed, a natural replacement of (\ref{LgaugeinvariantS}) is given by
\begin{equation}
\label{Lkineta}
\intS {\mbox{\boldmath{$\eta$}}}\, \L_{\rm kin} + \intSbar {{\mbox{\boldmath{$\etabar$}}}}\, {\overline
\L}_{\rm kin}
\end{equation}
with a dimensionless chiral
superfield 
$\mbox{\boldmath{$\eta$}}(x, \theta)$  
 and its complex conjugate 
$\mbox{\boldmath{$\etabar$}} (x, \thetabar)$, which are given by
\begin{equation}
\label{defeta}
{\mbox{\boldmath{$\eta$}}}(x, \theta) = \eta + \theta^ \alpha
 \chi_\alpha + \theta^2 f \ ,
 \qquad 
{{\mbox{\boldmath{$\etabar$}}} }(x, \thetabar) = \etabar + \theta_\alphadot \chibar ^
 \alphadot  + \thetabar^2 {\overline f} 
\end{equation}
in the
respective
chiral and
antichiral representation. The expression (\ref{Lkineta}) 
is supersymmetric and gauge invariant with
the transformations  (\ref{susyvector}), (\ref{gaugetr}). And (\ref{Lkineta}) contains
the gauge invariant expression (\ref{LgaugeinvariantS})
if we  identify the square of the local coupling 
with the
 inverse of the real part of $\eta$ and $\etabar$:
\begin{equation}
\label{loccoupl}
\frac{1}{e^2(x)} = {\eta (x) + \etabar(x)}.
\end{equation}



The relation (\ref{loccoupl}) specifies $e(x)$ as the lowest component
of a constrained real vector
multiplet
 $E (x, \theta, \thetabar )$:
\begin{eqnarray}
\label{E2def}
E (x, \theta, \thetabar ) & = & ({\mbox{\boldmath{$\eta$}}}(x, \theta,
\thetabar ) +{\mbox{\boldmath{$\etabar$}}}(x, \theta, \thetabar ))^ 
{-\frac 12} 
\end{eqnarray} 
(here ${\mbox{\boldmath{$\eta$}}}$ and ${\mbox{\boldmath{$\etabar$}}}$
have to be taken in the real representation),
and thus determines the supersymmetry transformations of the
local
gauge coupling as
\begin{equation}
\delta_\alpha e = -\frac 12 e^ 3 \chi_\alpha \qquad
\bar\delta_\alphadot e = -\frac 12 e^ 3 \chibar_\alphadot\ .
\end{equation}

With this identification  we find the following explicit expression
for
the gauge invariant and supersymmetric
action  (\ref{Lkineta}):
\equa{
& \Ga_{\rm kin} &  = 
\intd \Bigl(& - \frac 1{4e^2}F^ {\mn}(eA) F_{\mn} (eA) + \frac i2 (
\lambda \sigma \partial \lambdabar - \partial \lambda \sigma
\lambdabar)+\frac 1 8 D^2 \nonumber \\
&  &  + & \frac i 2 (\eta - \etabar ) \partial_\mu 
(e ^2 \lambda \sigma^ \mu \lambdabar)
- \frac i 8 (\eta - \etabar) \epsilon^{\mu \nu \rho\sigma} F_{\mn}(eA)
F_{\rho\sigma }(eA) \nonumber \\
 & & + &\frac i 4 e ( \chi \sigma^ {\mn} \lambda - \lambdabar {\bar
\sigma}^ {\mn} \chibar) F_{\mn}(eA)+ \frac i4e ^2 (\chi \lambda - \chibar
\lambdabar ) D \nonumber \\
 & & - & \frac 12 e ^2 f \lambda \lambda - \frac 12  e ^2 {\overline f}
\lambdabar \lambdabar \Bigr).
\label{Gakinext}
}

The complete action of SQED with local coupling is composed of
$\Ga_{\rm kin}$
(\ref{Gakinext}), the matter action $\Ga_{\rm matter}$ with the
Lagrangian (\ref{Lmatterex}) and the supersymmetric mass term 
$\Ga_{\rm mass}$ (see (\ref{Gasqed}) with (\ref{Lmass})).
It is supersymmetric under the transformations (\ref{susyvector}),
(\ref{susymatter}) and under transformations of the gauge coupling and
its superpartners
according to its definition (\ref{E2def}). It is remarkable that the
modifications of the local gauge coupling concern only the kinetic part
of the action, which includes now the chiral multiplet ${\cal L}_{\rm kin}$ and its
complex conjugate. The matter and mass terms remain unaffected
by the superfield extension of the coupling.


For the derivation of non-renormalization theorems it is important 
to note that the classical action depends on the
parity odd 
external field $\eta -\etabar$ only via a total derivative:
\begin{equation}
\label{holomorph}
\intd \Bigl( \frac {\delta} {\delta {\eta}} -  \frac {\delta}
 {\delta { \etabar}}\Bigr) \Ga_{\rm cl} = 0 .
\end{equation}
This identity constitutes a Ward identity expressing
 that the local coupling is the lowest component of a constrained
 real  superfield as defined in (\ref{E2def}), and it states 
the absence of purely chiral or
antichiral interactions in SQED.

\subsection{Chiral vertices}

It is  well-known that in the Wess--Zumino gauge the
non-renormalization of chiral vertices  is not manifest. It will turn
out that there is an underlying non-renormalization theorem, but the
non-renormalization properties are superposed by gauge-dependent field
renormalizations of the matter fields. In order to disentangle the
effect of the unphysical field renormalization from the
non-renormalization properties, it is useful to introduce a second
chiral vertex into the model. This can be done easily using 
a further    external  chiral and an antichiral field multiplet with
dimension 1:
\begin{equation}
\label{qdef}
{\mathbf q} =  q + \theta^\alpha q_\alpha + \theta^2 q_F\ ,  \qquad {\mathbf
\qbar} ={\overline q}  + \thetabar_\alphadot {\overline q}^\alphadot +
\thetabar^2 {\overline q}_F \ .
\end{equation}
 It can be coupled to the mass term as follows:
\begin{eqnarray}
\Ga_{\rm q} & = & 
\intd \Bigl( q  L_{\rm mass} + \bar q  \bar L_{\rm
mass} \nonumber \\
&  & \phantom{\intd} - \frac 1 {\sqrt 2}\bigl( q^ \alpha (\psi_{L\alpha} \varphi_R + 
\psi_{R\alpha} \varphi_L) +  \qbar_ \alphadot (\psibar_{L}^{\alphadot}
\phibar _R + 
\psibar_{R}^{\alphadot} \phibar_L)\bigr) \nonumber \\
& & \phantom{\intd} + q_F \varphi_L \varphi_R + \bar q _F \phibar_L
\phibar_R \Bigr) \ .
\label{Gaq} 
\end{eqnarray}
The action of SQED with local coupling being enlarged by $\Ga_{\rm q}$,
it  contains
 two chiral vertex functions $\Ga_{\psi\psi}$ and $\Ga_{q\psi\psi}$,
and we will obtain non-trivial non-renor\-mali\-zation theorems relating
the divergences appearing in $\Ga_{q\psi\psi}$ and $\Ga_{\psi\psi}$.

Furthermore it will turn out that the $\mathbf q $-multiplet and its
complex conjugate
 take an important role when   we gauge softly broken
axial symmetry and when we construct the Callan--Symanzik equation.

\section{Quantization of SQED  with local  coupling }
\subsection{The Slavnov--Taylor identity}

For quantizing SQED with local
gauge coupling  in the Wess--Zumino gauge
all  symmetries, gauge symmetry, supersymmetry and translational
symmetry, are included  into an enlarged
Slavnov--Taylor identity \cite{White92a, MPW96a}.
As in ordinary SQED \cite{HKS99},
all transformation parameters are replaced by the respective
ghosts. So we introduce the  
Faddeev--Popov ghost $c(x)$ for gauge transformations,  and 
the  space-time independent supersymmetry and translation
ghosts $\epsilon^\alpha,\epsilonbar^\alphadot$ and $\omega^\nu$.

Then we eliminate  the auxiliary fields 
 $D$ and $F_L, F_R, \bar F_L , \bar F_R$  by their equations of motions,
\begin{eqnarray}
\label{FDexpr}
D&  =  & -ie^2 (\chi \lambda - \chibar \lambdabar) -2  e  Q_L (\varphi_L
\phibar _L -\varphi_R
\phibar _R ) \ ,\nonumber \\
{\overline F}_L & = & - (q +m ) \varphi_R \ ,\qquad \quad
{ F}_L  =  - (\qbar +m ) \phibar_R \ ,\nonumber \\
  {\overline F}_R & = & - (q +m ) \varphi_L \ ,
\quad \qquad { F}_R  =  - (\qbar +m ) \phibar_L \ ,
\end{eqnarray}
and get an action in terms of physical fields.

All symmetries and the structure constants of the algebra are
summarized in the BRS transformations listed in appendix C.
After the elimination of the auxiliary fields the
BRS transformations are nilpotent 
only up to equations of
motions.

Owing to the nilpotency  of the BRS transformations it is straightforward  to 
add a BRS invariant gauge fixing and Faddeev--Popov field part to the
action. 
 We
choose it in such a way that the gauge fixing is stable under
renormalization and results in the usual gauge fixing for constant
coupling. Using the auxiliary field $B$, the Faddeev-Popov anti-ghost
$\cbar$ and the local gauge parameter $\xi(x) + \xi$ it reads:
\begin{equation}
\label{gaugefixing}
\Ga_{\rm g.f}+ \Ga_{\phi\pi} = 
{\mathrm s} \intd (\frac 12(\xi(x) + \xi) \cbar B + \frac 1e \cbar
\partial (eA) )\ .
\end{equation}
Working this out
with the BRS transformations of the appendix we get:
\begin{eqnarray}
\Gamma_{\rm g.f.}+ \Ga_{\phi\pi} & = & 
 \intd \Bigl( \frac 1 e B\partial^\mu(e A_\mu) + \frac 12 (\xi(x)+ \xi)B^2
- \frac 1 e \bar{c}\Box (ec) + \frac 12 \chi_\xi \cbar B
\nonumber\\
&&{}\quad
- \frac 1 e \cbar\partial^\mu(i\epsilon\sigma_\mu e\lambdabar 
                     -ie \lambda\sigma_\mu\epsilonbar) 
+ ( \xi(x) + \xi) i \epsilon\sigma^\nu\epsilonbar
  (\partial_\nu\cbar)\cbar \Bigr) \ .
\label{GaugeFixingTerm}
\end{eqnarray}
We want to note that the local gauge fixing
field $\xi(x)$ is introduced for later use in the Callan--Symanzik equation.
For a space-time dependent gauge fixing parameter $\xi(x)$
supersymmetry
 and translational invariance enforces  to
introduce its BRS partner $\chi_\xi(x)$.
 BRS variations of the gauge parameter have already been introduced
in ordinary gauge theories for controlling gauge parameter dependence
of Green functions \cite{PISI85} and are an important tool
for identifying gauge invariant quantities \cite{HAKR97}.


For the formulation of the Slavnov--Taylor identity one introduces the
external field part, which also contains the bilinear part for absorbing
the equations of motion terms violating the nilpotency of BRS
transformations
\cite{BV}:
\begin{eqnarray}
\Gamma_{\rm ext} & = & \intd \Bigl(Y_\lambda^\alpha s\lambda_\alpha
+ Y_\lambdabar{}_\alphadot s\lambdabar^\alphadot
\nonumber\\&&{}\hfill
+ Y_{\varphi_L} s\varphi_L + Y_{\phibar_L} s\phibar_L
+ Y_{\psi_L}^\alpha s\psi_L{}_\alpha
+ Y_{\psibar_L}{}_\alphadot s\psibar_L^\alphadot
+ ({}_{L\to R}) \nonumber\\&&{ }\hfill
+  \frac 12(Y_\lambda \epsilon-\epsilonbar
                     Y_\lambdabar)^ 2
                   -2(Y_{\psi_L} \epsilon)(\epsilonbar  Y_{\psibar_L})
                   -2(Y_{\psi_R} \epsilon)(\epsilonbar  Y_{\psibar_R})\Bigr)
 \  .
\end{eqnarray}
It coincides in its structure with the one of
ordinary SQED \cite{HKS99}.
The complete classical action
\begin{equation}
\Gamma_{\rm cl}  = \Gamma_{\rm SQED} + \Ga_{\rm g.f.} +
\Ga_{\phi\pi} +
\Gamma_{\rm ext} 
 \, , 
\end{equation}
 satisfies the
Slavnov--Taylor
identity:
\begin{equation}\label{ST}
{\cal S} (\Ga_{\rm cl}) = 0\ .
\end{equation}
 The Slavnov--Taylor operator acting on a general
functional ${\cal F}$  is defined as
\begin{eqnarray}
{\cal S}({\cal F}) & = & 
\intd\Bigl(s A^\mu \dF{A^\mu}
+ \dF{Y_\lambda{}_\alpha}\dF{\lambda^\alpha}
+ \dF{Y_\lambdabar^\alphadot}\dF{\lambdabar_\alphadot}
 \nonumber\\&&{}\quad
+   s c \dF{c}         + s B \dF{B} + s\bar{c} \dF{\bar{c}} + s \xi \dF{\xi}
+  s\chi_\xi \dF{\chi_\xi} 
\nonumber\\&&{}\quad
+ \dF{Y_{\varphi_L}}\dF{\varphi_L}
+ \dF{Y_{\phibar_L}}\dF{\phibar_L}
+ \dF{Y_{\psi_L{}_\alpha}}\dF{\psi_L^\alpha}
+ \dF{Y_{\psibar_L}^\alphadot}\dF{\psibar_L{}_\alphadot}
+(_{L\to R})\Bigr)
\nonumber\\&&{} \quad
+ s\eta^i\frac{\delta{\cal F}}{\delta\eta^ i}
+ s\etabar^i\frac{\delta{\cal F}}{\delta\etabar^ i}
+ s q^i\frac{\delta{\cal F}}{\delta q^ i}
+ s\qbar^i\frac{\delta{\cal F}}{\delta\qbar^ i}
+ \Bigr)
+ s\omega^\nu \frac{\partial{\cal F}}{\partial\omega^\nu} \ .
\label{STOperator}
\end{eqnarray}
In comparison to usual SQED the Slavnov--Taylor operator contains in
addition
the BRS transformations  of the chiral
 multiplet ${\eta^ i}  =  (\eta, \chi^\alpha,  f)$ and its complex
conjugate antichiral multiplet $\bar \eta^ i$, which define
 the local coupling (\ref{loccoupl}), the BRS transformation of the
chiral multiplet ${q^ i}  =  (q, q^\alpha,  q_F)$ and its complex
conjugate $\qbar^ i$ (\ref{qdef}) and the BRS transformations of the
gauge parameter doublet.
 
The full Slavnov--Taylor operator and its linearized version have the
nilpotency property 
\begin{eqnarray}
\label{nilpotency}
s_{\cal F} S({\cal F}) & = & 0
\end{eqnarray}
if the functional ${\cal F}$ satisfies the linear identity
\begin{eqnarray}
i\epsilon\sigma^\mu\frac{\delta\cal F}{\delta  Y_\lambdabar}
+i\frac{\delta\cal F}{\delta Y_\lambda}\sigma^\mu\epsilonbar  & = &
 \bigl(\frac 12 e^ 2 
(\epsilon \chi-\chibar \epsilonbar)
-i\omega^\nu\partial_\nu\bigr)(i\epsilon\sigma^\mu\lambdabar
                        -i\lambda\sigma^\mu\epsilonbar) \nonumber \\
 & & \; +\,2i\epsilon\sigma_\nu\epsilonbar \frac 1 e F^{\nu\mu} (eA)\ .
\label{EquivNilpotency}
\end{eqnarray}
This linear identity  guarantees the 
nilpotency properties  on the linear transformations of the
 photon field $A^\mu$:
 $s_{\cal F}^2 A^\mu=0$. 
Eq.~(\ref{EquivNilpotency}) is satisfied in particular by $\Gamma_{\rm
  cl}$ and can be maintained also in the course
of
renormalization.

\subsection{Renormalization}

 The local gauge coupling and its superpartners
 (\ref{E2def}) are considered as  external fields 
which appear  in the same
way as ordinary
external fields in the generating functional of 1PI Green functions
$\Ga$. In ordinary SQED the coupling constant is the perturbative
expansion parameter. By a simple consideration of diagrams it can be seen
that this property is true also for the 
 local coupling $e(x)$.  Indeed, the number of local
couplings
 appearing in a specific diagram
is related to the loop order $l$ 
by a 
topological formula:
\begin{equation}
\label{topfor}
N_{e(x)} = N_{\rm amp. legs} + N_Y+ 2N_f+
2N_\chi + 2 N_{\eta - \etabar} 
+ 2(l-1)\ .
\end{equation}
Here  $N_{\rm amp. legs} $ counts the number of external
amputated legs with propagating fields ($A^ \mu, \lambda, \varphi_A,
\psi_A, c, \cbar  $ and the respective
complex conjugate fields), $N_Y$ gives the number of BRS insertions,
counted by the number  of differentiations with respect the the
external fields $Y_\phi$. 
 $N_f$, $N_\chi$ and $N_{\eta - \etabar } $ gives the number of 
insertions 
corresponding to  the respective external fields. These  are the
$\eta - \etabar$ and 
higher  components of the chiral and antichiral
multiplets defining the local coupling. 
By the existence of a topological formula
 the local 
coupling is    distinguished from  
ordinary dimensionless external fields like the spurion fields
\cite{GIGR82, HKS00},
 which can appear in arbitrary
orders in the Green functions of higher order perturbation theory.

We are  able to renormalize the Green functions of SQED 
with local coupling as a
simple extension of usual SQED. The Slavnov--Taylor identity of SQED with local
couplings
is --- as for usual SQED --- not anomalous, and one is able to
  establish the Slavnov--Taylor identity 
\begin{equation}
\label{STidentity}
{\cal S} (\Ga)= 0
\end{equation}
and the linear equation
(\ref{EquivNilpotency}) to all orders.
Furthermore we require the linear gauge fixing function
(see (\ref{GaugeFixingTerm})) as normalization for the $B$-field
and the linear ghost equations,
\begin{equation}
\label{ghostequ}
\frac{\delta \Ga} {\delta c} =
\frac{\delta \Gacl} {\delta c}  \qquad \mbox{\rm and} \qquad 
\frac{\partial \Ga} {\partial \omega^ \mu} =
\frac{\partial \Gacl} {\partial \omega^ \mu} \,
\end{equation}
as normalization conditions for the ghosts.
As in ordinary SQED  
the abelian gauge Ward identity is derived from the consistency
equation of the ghost equation with the Slavnov--Taylor identity
\cite{HKS99, MPW96a}. With local gauge coupling it takes  the form
\begin{eqnarray}
\Bigl( {\bf w}^ {\rm em} - \partial^\mu \bigr( \frac 1e
\frac {\delta} {\delta {A^\mu}}\bigl)\Bigr)\Ga &  = & \Box(\frac 1e B)
 +{\cal O}(\omega) \ .
\label{WIem}
\end{eqnarray}
In the adiabatic 
limit the  real  superfield of the coupling becomes a constant
\begin{equation}
\label{limitSQED}
E(x, \theta , \thetabar) \to e = \mbox{const},
\end{equation}
and one recovers the 1PI Green functions and Ward identities of
ordinary SQED as defined in \cite{HKS99},
\begin{equation}
\label{SQEDlimit}
\lim_{E \to e} \Ga = \Ga^{\rm SQED}.
\end{equation}

As a new ingredient one requires in addition to the above symmetries
the identity 
 (\ref{holomorph}):
\begin{equation}
\label{holomorphGa}
\intd \Bigl(\frac {\delta } {\delta \eta}  - \frac {\delta} {\delta \etabar}
\Bigr) \Ga = 0\ .
\end{equation}
It is valid in the classical approximation and can be extended
to higher orders of perturbation theory.
In combination with supersymmetry, eq.~(\ref{holomorphGa}) is the
crucial identity
 for the construction of non-renormalization
theorems in SQED.

\section{Improved renormalization behaviour}

\subsection{Invariant counterterms}

The simplest, but somewhat indirect way to derive
 the improved renormalization behavior of 1PI Green functions can be carried
 out by an investigation
of symmetric counterterms: Absence of symmetric counterterms to 
 Green functions means
 that the corresponding Green functions are related to non-local
 expressions, which  cannot appear with independent renormalizations.
In this section we construct  the invariant counterterms of higher
 orders and find that  invariant counterterms to the
 photon self energy from two-loop order onwards and  independent
 counterterms to the chiral vertices are absent. The algebraic reason for the
 absence of these counterterms is the identity (\ref{holomorph}),
 which characterizes the local gauge 
coupling as a constrained  real superfield.

Due to the  nilpotency properties of the  Slavnov--Taylor
operator the 
 invariant counterterms  in loop order $l$
are restricted by the  symmetries of the classical action, i.e.
\begin{equation}
\label{ctconstraints}
s_{\Ga_{\rm cl}} \Ga^ {(l)}_{\rm ct,inv}  =  0 \qquad \hbox{and} \qquad
\intd \Bigl(\frac {\delta } {\delta \eta}  - \frac {\delta} {\delta \etabar}
\Bigr) \Ga^ {(l)}_{\rm ct,inv}  =  0 \ ,
\end{equation}
and $\Ga_{\rm ct, inv}$ is invariant under the discrete symmetries C,
P and R-parity (see
  appendix B).
A further constraint on the counterterms is the topological formula
(\ref{topfor}), which
determines the order in the local coupling.

All symmetric
counterterms can be algebraically classified  as 
 non-BRS variations and BRS variations. Non-BRS variations comprise
the renormalization of physical parameters, whereas BRS variations
are connected with unphysical field renormalizations. In contrast
to the non-BRS variations, BRS variations receive gauge-dependent coefficients, as can
be easily derived using the BRS varying gauge parameter
(cf.~\cite{PISI85} for a detailed discussion).

We start the construction of invariant counterterms with the physically
relevant ones, the non-BRS variations. 
 The simplest and most obvious way to construct them
 is given by their superspace formulation. Therefore we use  the multiplet
structure of the gauge invariant Lagrangians of section 2.1 and
eliminate the auxiliary fields in the end. Indeed, since the non-BRS
variations  are
gauge-independent, 
they appear in the same way
 in the supersymmetric gauge and are not affected by the
elimination of the auxiliary fields.

Particular supersymmetric
and gauge invariant counterterms are the counterterms to the chiral vertices
\begin{eqnarray}
\label{ctmass}
\intS {\mbox{\boldmath{$\eta$}}}^{-l} \L_{\rm mass} +
\intSbar {\mbox{\boldmath{$\eta$}}}^ {-l} \bar \L_{\rm mass} ,
\quad \intS {\mbox{\boldmath{$\eta$}}}^{-l}\mathbf q \L_{\rm mass} +
\intSbar {\mbox{\boldmath{$\eta$}}}^ {-l}\bar {\mathbf q} \bar \L_{\rm
mass} ,
\end{eqnarray} 
and the counterterm to the kinetic term of the photon multiplet
\begin{equation}
\label{ctkin}
\intS {\mbox{\boldmath{$\eta$}}}^{-l+1} \L_{\rm kin} +
\intSbar {\mbox{\boldmath{$\etabar$}}}^ {-l+1} \bar \L_{\rm kin}\ .
\end{equation}
Here the dependence on the loop order $l$ is governed by the
topological formula (\ref{topfor}).
However, the symmetric counterterms have to satisfy the second
identity in (\ref{ctconstraints}), too. Indeed, (\ref{ctmass}) does not
satisfy (\ref{ctconstraints}) for $l \neq 0$, and (\ref{ctkin}) does not
satisfy (\ref{ctconstraints}) for $l \geq  2$. 
Accordingly, loop diagrams to these non-symmetric
counterterms  are not present
since the classical action only depends  the constrained real
superfield ${\mbox{\boldmath{$\eta$}}} +
{\mbox{\boldmath{$\etabar$}}}$ but not on the single chiral or
antichiral fields. 
Thus, the only of the above
counterterms that is in agreement with both identities in (\ref{ctconstraints}) is the 1-loop counterterm to the
kinetic terms of the photon multiplet:
\begin{equation}
\Ga^ {(1)}_{\rm ct,kin} =  \intd \Bigl(
F^ {\mn} (eA)F_{\mn}(eA) + e^ 2 i \lambda \sigma^ \mu \partial \lambdabar
+  \frac 18 e^2 D^ 2\Bigr).
\end{equation}

 There exists  a
 further   field monomial which is a non-BRS variation:
\begin{equation}
\label{ctmatter}
\Ga^ {(l)}_{\rm ct, matter}= \frac 1{16}\intV E^ {2l} \L_{\rm matter}\ ,
\end{equation}
where $E^{2l}$ and $\L_{\rm matter}$ are the real  multiplets
corresponding to $e^{2l}$ and $L_{\rm matter}$, respectively:
\begin{eqnarray}
\label{E2l}
E^{2 l}(x, \theta, \thetabar )  & = & \bigl({\mbox{\boldmath{$\eta$}}}(x,
\theta, \thetabar ) +{\mbox{\boldmath{$\etabar$}}}(x, \theta,
\thetabar )\bigr)^ 
{-l} 
\end{eqnarray}
and 
\equa{\label{Lmattervek}
&{\L}_{{\rm matter},A} (x,\theta,\thetabar)  = \nonumber \\
& \qquad\phibar_A \varphi_A + \theta \sqrt 2 \psi_A \phibar_A + \thetabar 
\sqrt 2 \psibar_A
\varphi_A
+ \theta ^ 2 F_A \phibar_A + \thetabar^ 2 {\bar F }_A \varphi_A\nonumber \\
& \quad
 +\; \theta \sigma^ \mu \thetabar (i \varphi_A D_\mu \phibar_A - i \phibar_A
D_{\mu} \varphi_A +  \psi_A \sigma^ \mu \psibar_A ) \nonumber \\
& \quad +\;  \thetabar^ 2 \theta \bigl({ \sigma}^ \mu 
\frac i {\sqrt 2}( \psibar_A D_{\mu} \varphi_A
-  D_{\mu } \psibar_A  \varphi_A) 
- 2ie Q_A  \lambda \varphi_A \phibar _A  + {\sqrt 2 } \psi {\overline F}_A
\bigr) 
 \nonumber \\
& \quad+ \; \theta^ 2 \thetabar \bigl(\bar{ \sigma}^ \mu 
\frac i {\sqrt 2}( \psi_A D_{\mu} \phibar_A
-  D_{\mu } \psi_A  \phibar_A) 
+ 2ie Q_A  \lambdabar \varphi_A \phibar _A  + {\sqrt 2 } \psibar F_A\bigr) 
 \nonumber \\
&\quad + \;\theta^ 2 \thetabar ^2  L_{{\rm matter}, A} \ .
}
The two counterterms $\Ga_{\rm ct, kin}$ and $\Ga_{\rm ct,matter}$ 
are the only gauge-independent counterterms, i.e.
\begin{equation}
\Ga^ {(l)}_{\rm ct, nonBRS} = 
 z_{\rm kin}^ {(1)} \Ga^ {(1)}_{\rm ct,  kin} \delta_{l1} +
z_{m}^ {(l)}\Ga^ {(l)}_{\rm ct, matter}  \ .
\end{equation}

Finally we have to eliminate
the auxiliary fields $D$ and $F$ from $\Ga^ {(1)}_{\rm ct,  kin}$  and
$\Ga^ {(l)}_{\rm ct, matter}  $ by their
equations of motion:
\begin{equation}
 \frac {\delta} {\delta D}( \Ga_{\rm cl}(D, F) +   \Ga^ {(l)}_{\rm ct, nonBRS}(D, F) ) =
0 
\end{equation}
and respective expressions for the $F$-fields. When we insert the result
into the action and  put together
all terms of loop order $l$, 
we obtain the $s_{\Gacl}$-invariant
counterterms without auxiliary fields.

The  explicit form of the corresponding symmetric counterterms can be
found in a simpler way by expressing the invariant counterterms
in terms of symmetric operators. For the counterterm (\ref{ctmatter}),
however, 
an invariant operator can only be constructed with the help of a
further axial vector multiplet (see (\ref{Dkin}) and
(\ref{MatterCt})). Nevertheless, it is possible and
sufficient for our purposes to consider the limit to constant coupling
here. 
For constant coupling, $\Ga^ {(1)}_{\rm
ct,kin}$
corresponds to the usual coupling renormalization
\begin{eqnarray}
\label{Dkine}
\lim_{E\to e}\Ga^ {(1)}_{\rm ct, kin} &= &-\frac 12 e^ 2  \Bigl(
e \partial_e  - N_A - N_c -  N_\lambda + N_{Y_\lambda} \\
& & \phantom{-\frac 12 e^ 2} + N_B +
N_{\cbar}
- 2 \xi \partial_\xi \Bigr) \Ga _{\rm cl}^ {\rm SQED}\ ,  \nonumber
\end{eqnarray}
and  
$\Ga_{\rm ct, matter}$ can be decomposed into a field and mass
renormalization:
\begin{equation}
\label{ctmattere}
\lim_{E \to e  } \Ga^ {(l)}_{\rm ct, matter}  = \frac 12 e^ {2l}
\bigl( (N_\varphi 
+ N_\psi - N_{Y_\psi} - N_{Y_\varphi} )
- 2  (N_{\mathbf q} + m\partial_m )\bigr)\Ga_{\rm cl} ^ {\rm SQED}\ .
\end{equation}
In (\ref{Dkine}) and (\ref{ctmattere}) the operators $N_\phi$  denote 
 the usual field counting operators including 
also the 
 complex conjugates for complex fields.
Eqs.~(\ref{Dkine}) and (\ref{ctmattere}) determine the corresponding 
invariant counterterms without auxiliary fields in the limit to
 constant coupling.

 The bilinear form of the Slavnov--Taylor
identity gives  rise to three kinds of unphysical counterterms
corresponding to field renormalizations of the
 matter fields:
\begin{eqnarray}
&&\psi_A \to z^ {(l)}_\psi e^ {2l} \psi_A\ , \qquad
\varphi_A \to z^ {(l)}_\varphi e ^{2l} \varphi_A\ , \nonumber \\
&& \psi_A \to z_{\psi\varphi}e^ {2l+2} \chi \varphi_A\ .
\end{eqnarray}
These  field renormalizations are
BRS variations, and the coefficients $z^ {(l)}$ are therefore gauge
dependent. In explicit 
calculations   
  they indeed appear  with non-vanishing coefficients in the
Wess--Zumino gauge (see e.g.\ \cite{HKS99}). The
invariant counterterms
corresponding to these field
renormalizations are best given in the form of a symmetric operator acting
on the classical action: 
\begin{eqnarray}
 \Ga^ {(l)}_{{\rm ct},\varphi} & = & s_{\Ga_{\rm cl}} \intd e^{2l} f^ {(l)}_{\varphi}(\tilde \xi)
\bigl(Y_{\varphi_L} \varphi_L +  Y_{\varphi_R} \varphi_R +
Y_{\phibar_L} \phibar_L +   Y_{\phibar_R} \phibar_R  \bigr) \nonumber \\
& = &  
\intd e^ {2l} f^ {(l)}_\varphi( \tilde \xi) \biggl(
\Bigl(\varphi_L \frac {\delta}{\delta \varphi_L } -
Y_{\varphi_L} \frac {\delta}{\delta Y_{\varphi_L }} + {\rm c.c.}\Bigr) + 
(_{L\to R})\biggr)\Ga_{\rm cl} \nonumber \\
& & + \intd s \bigl(e^{2l} 
  f^{(l)}_{\varphi}(\tilde \xi))\bigr)
\bigl(Y_{\varphi_L} \varphi_L +  Y_{\varphi_R} \varphi_R +
Y_{\phibar_L} \phibar_L +   Y_{\phibar_R} \phibar_R
\bigr)\biggr)\nonumber \\
& \equiv & {\cal N}^ {(l)}_\varphi \Gacl + \Delta^ {(l)}_{Y,\varphi}\ ,
\label{ctvarphi} 
\\
\Ga^ {(l)}_{{\rm ct},\psi} & = & s_{\Ga_{\rm cl}} \intd e^{2l} f^ {(l)}_{\psi}(\tilde \xi)
\bigl(\psi_L Y_{\psi_L}  +\psi_R  Y_{\psi_R}  +
\psibar_L Y_{\psibar_L}  + \psibar_R  Y_{\psibar_R}   \bigr) \nonumber \\
& = &  
\intd e^ {2l} f^ {(l)}_\psi( \tilde \xi) \biggl(
\Bigl(\psi^ \alpha_L \frac {\delta}{\delta \psi^ \alpha_L } -
Y^ \alpha_{\psi_L} \frac {\delta}{\delta Y^ \alpha_{\psi_L }} +
 {\rm c.c.}\Bigr) + 
(_{L\to R})\biggr)\Ga_{\rm cl} \nonumber \\
& & + \intd s \bigl(e^{2l} 
  f^ {(l)}_{\psi}(\tilde \xi))\bigr)
\bigl( \psi_L Y_{\psi_L}  +\psi_R  Y_{\psi_R}  +
\psibar_L Y_{\psibar_L}  + \psibar_R  Y_{\psibar_R}   \bigr)\biggr)
\nonumber\\
& \equiv & {\cal N}^ {(l)}_\psi \Gacl + \Delta^ {(l)}_{Y,\psi}\ ,
\label{ctpsi} 
\\
 \Ga^ {(l)}_{{\rm ct},\psi\varphi} & = &
 s_{\Ga_{\rm cl}} \intd \sqrt 2  f^ {(l)}_{\psi\varphi}(\tilde \xi)
\bigl(\chi^ {E^ {(2l)}} \varphi_L Y_{\psi_L}  + \chibar^  {E^ {(2l)}}
\phibar_L Y_{\psibar_L} + (_{L\to R})\bigr)  \nonumber\\ 
 \nonumber \\
& = & - l  \intd \sqrt 2 e^{ 2(l+1)}f^ {(l)}_{\psi\varphi}(\tilde \xi) \times
\nonumber \\
& & \phantom{ -l \intd \sqrt 2 e^ {2}}
 \biggl(\Bigl(\chi^ \alpha \bigr(\varphi_L
\frac {\delta }{ \delta \psi^ \alpha_L}
-  Y_{\psi_L \alpha}
\frac {\delta } {\delta Y_{\varphi}}\bigl) + {\rm c.c.}\Bigr) +
 (_{L\to R}) \biggr)
\Ga_{\rm cl}
\nonumber \\
& & -l \intd s \bigl(\sqrt 2 e^{2(l+1)} 
  f^ {(l)}_{\psi\varphi}(\tilde \xi)\bigr)
\bigl(\chi \varphi_L Y_{\psi_L} +
\chibar \phibar_L Y_{\psibar_L} + (_{L\to R})
\bigr) \nonumber \\
& & -l \intd  \sqrt 2 e^{2(l+1)} 
  f^ {(l)}_{\psi\varphi}(\tilde \xi)
\bigl(s (\chi) \varphi_L Y_{\psi_L} +
s (\chibar) \phibar_L Y_{\psibar_L} + (_{L\to R})
\bigr) \nonumber \\
& \equiv & {\cal N}^ {(l)}_{\psi\varphi} \Gacl + \Delta^
{(l)}_{Y,\psi\varphi} .
\label{ctpsivarphi}
\end{eqnarray}

\subsection{Non-renormalization theorems}

The complete action of symmetric counterterms $\Ga_{\rm ct, inv}$
in loop order $l$ is a linear combination of the single 
BRS invariant parts constructed above:
\begin{eqnarray}
\label{Gactinv}
\Ga^ {(l)}_{\rm ct, inv} & = &
 z_{\rm kin}^ {(1)} \Ga^ {(1)}_{\rm ct,  kin} \delta_{l1} +
z_{m}^ {(l)}\Ga^ {(l)}_{\rm ct, matter}  \nonumber \\
& +&
z_\varphi^ {(l)}\Ga^ {(l)}_{{\rm ct}, \varphi}(\xi) +
z_\psi^ {(l)}\Ga^ {(l)}_{{\rm ct}, \psi}(\xi) +
z_{\psi\varphi}^ {(l)}\Ga^ {(l)}_{{\rm ct}, \psi\varphi} (\xi)\ .
\end{eqnarray}
Its independent coefficients are the 1-loop coefficient of
the kinetic term of the photon and photino $z^ {(1)}_{\rm kin}$, the
coefficients of the matter part $z_{m}^ {(l)}$, and the coefficients
of the field renormalizations $z_\varphi^ {(l)}, z_\psi^ {(l)}$ and
$z_{\psi\varphi}^ {(l)}$.

Comparing this result with the counterterms of usual SQED with
constant
coupling yields the non-renormalization theorems.

Using eqs.~(\ref{Dkine}) and (\ref{ctmattere}) the symmetric
counterterms
of SQED are determined 
 by the following expression:
\begin{eqnarray}
\label{ctidentification}
\lim_{E \to e  } \Ga^ {(l)}_{\rm ct, inv} & = & 
\Bigr(\delta Z_e^ {(1)}   \bigl(
e \partial_e  - N_A - N_c -  N_\lambda + N_{Y_\lambda} + N_B +
N_{\cbar}
- 2 \xi \partial_\xi \bigr) \\
& & { }+ \delta Z_m^ {(l)} (N_q + N_{q ^\alpha} + N_{q_F} + m\partial_m )
\nonumber \\
& &{  } +  \delta Z^{(l)} _\varphi (N_\varphi - N_{Y_\varphi})
+ \delta Z^ {(l)}_\psi (N_\psi - N_{Y_\psi}) 
\Bigr)\Ga^ {\rm SQED}_{\rm
cl}\ . \nonumber 
\end{eqnarray}
Here  we have defined the $z$-factors as power series in the coupling:
\begin{eqnarray}
\label{zfactor}
\delta Z^ {(1)}_e & = &  - \frac 12 z^ {(1)}_{\rm kin} e ^ 2\ ,
\qquad \delta Z_ m^ {(l)} = - z_m^ {(l)} e  ^{2l}\ , \\
\delta Z^ {(l)}_\varphi & = &e^{ 2 l}( z^ {(l)}_\varphi  f^
{(l)}_\varphi(\xi) + \frac 12 z_m^ {(l)})\ ,\nonumber \\
\delta Z^ {(l)}_\psi & = e^{ 2 l} (
& z^ {(l)}_\psi f^ {(l)}_\psi(\xi) + \frac 12 z_m^{(l)}) \ .
\nonumber 
\end{eqnarray}

These restrictions on the counterterms constitute the
non-renormalization theorem of chiral vertex functions and the generalized
non-renormalization theorem of the photon self energy from two-loop order
onwards. Non-renormalization of chiral vertex functions is expressed by
the common $z$-factor $\delta
Z_m$ of $q$-field and mass renormalization. 
Hence, in the Wess--Zumino gauge the non-renormalization of chiral vertices
 is hidden by the appearance of individual gauge-dependent field
 renormalizations to all matter fields.

As can be immediately seen from (\ref{zfactor}), 
the non-renormalization of chiral vertices would be  manifest if  the gauge
dependent field renormalizations $ z_\varphi$ and $z_\psi$  vanished
and $\delta Z_\varphi$ and $\delta 
Z_\psi$
were related to $\delta Z_m$. But this is the case only in the
manifestly supersymmetric gauge.

The absence of  counterterms to the photon self energy from two-loop
order onwards is a remarkable result of the construction with a local
coupling. It implies that the photon self energy is related to non-local Green
functions in the present construction.
In section 8 the relation of this result to the well-known restrictions on the gauge
$\beta$-function  will be worked out.


\section{Non-renormalization theorems in explicit \\ expressions}

In this section we explicitly work out the non-renormalization 
theorem of chiral
vertices in the example of the $\Ga_{\psi \psi}$  and $\Ga_{q\psi
\psi}$ vertices and the non-renormalization of the photon self energy
in loop order $l\geq 2$.

The procedure we apply is similar to the one which has already been applied
in the context of the Wess--Zumino model \cite{FLKR00}: Due to local
couplings one is able to  extract a supersymmetry transformation
 $\bar \delta_\alphadot$ from
 the  vertex functions of ordinary SQED  in loop orders $l \geq 1$.
 Technically such an extraction is performed by
working out $\frac{\delta S(\Gamma)}{\delta\epsilonbar^
  \alphadot\delta\chibar_\alphadot(x)}$. The result yields
\begin{equation}
\label{STchibar}
\lim_{E\to e}  \intd  \Bigl( 2 \frac {\delta}{\delta \etabar}   +   e
^2 A^ \mu \frac 
{\delta}{\delta A^ \mu}\Bigr) \Ga =  \lim_{E\to e}\frac {\partial}
{\partial \epsilonbar ^ 
\alphadot}\Bigl(
s_\Ga  \bigl(
 -  \intd \frac {\delta} {\delta \chibar_\alphadot(x)} \Ga \bigr)\Bigr).
\end{equation}

For evaluating the l.h.s.~of (\ref{STchibar})  
  the identity
\begin{equation}
\label{holomorph2}
\intd\Bigl(\frac {\delta}{\delta \eta} -\frac {\delta}{\delta
\etabar}
\Bigr) \Ga = 0
\end{equation}
becomes relevant. Using (\ref{loccoupl}) it enables us to relate the integrated field
 differentiation with respect to $\etabar$ immediately to a
 differentiation with respect to the coupling:
\begin{equation}
\label{eetabar}
\intd \frac {\delta}{\delta \etabar} \Ga  = 
\intd \frac {\delta}{\delta \eta} \Ga  = 
- \frac 12 \intd e^ 3(x) \frac {\delta} {\delta e} \Ga \ .
\end{equation}
Now we can use (\ref{SQEDlimit}) in (\ref{STchibar}) and obtain
\begin{equation}
\label{STchibar2}
\Bigl(- e^ 3 \partial_e   +   e ^2 \intd A^ \mu \frac 
{\delta}{\delta A^ \mu}\Bigr) \Ga^ {\rm SQED} = \lim_{E\to e}\frac {\partial}
{\partial \epsilonbar ^ 
\alphadot}\Bigl(
s_\Ga  \bigl(-
   \intd \frac {\delta} {\delta \chibar_\alphadot(x)} \Ga
   \bigr)\Bigr)\ .
\end{equation}
This is the desired
relation between SQED Green functions and the Green functions with a
vertex insertion of the spinor component of the chiral multiplet
 $\bar {\cal L}_{\rm kin}$.
In the following we evaluate the right-hand-side of
 eq.~(\ref{STchibar2}) and obtain the  non-local expressions for the 
  Green functions 
we are looking for.

\subsection{The chiral vertex functions}
 
First  we  derive explicit expressions for the chiral
vertex functions $\Ga_{\psi_L\psi_R}$ and $\Ga_{q \psi_L \psi_R}$.
These chiral vertex functions are 
superficially divergent in the Wess--Zumino gauge. But from the
expression
$\Ga_{\rm ct,inv}$ (\ref{Gactinv})
 the common origin of their divergences can be identified as the gauge
dependent field redefinition $\Ga_{{\rm ct}, \psi}$, so both
divergences arise with the same
coefficient $z_\psi f_\psi(\xi)$.
This structure can be made explicit by evaluating the identity (\ref{STchibar2}) for the
vertex functions $\Ga_{\psi_L\psi_R}$ and $\Ga_{q \psi_L \psi_R}$.

Differentiating  the identity (\ref{STchibar2})  
 with
respect to 
$ \psi_L^ \beta (x)$  and $\psi_R^ \gamma (y)$ yields
\begin{eqnarray}
\label{Gapsi}
& & e^ 3 \partial_e \Ga _{\psi_L^ \beta \psi_R^ \gamma} (p,-p) = 
\\
& &     \Ga_{\epsilonbar ^ {\betadot}\psi^\beta_L   Y_{\phibar_R}}(p,-p)
\Ga_{\chibar_{\betadot} \psi^\gamma_R   {\phibar_R}} (-p ,p)
- \Ga_{\epsilonbar ^ {\betadot}\psi^\gamma_R   Y_{\phibar_L}}(-p,p)
\Ga_{\chibar_{\betadot} \psi^\beta_L   {\phibar_L}}(p,-p) \nonumber \\
&- & 
 \Ga_{ \epsilonbar ^ {\betadot}\chibar_\betadot
\psi^\beta_L 
Y_{\psibar_R}^ \alphadot }(p,-p)  
\Ga_{\psi^\gamma_R   {\psibar_{R\alphadot}}}(-p,p)
+ \Ga_{\epsilonbar ^ {\betadot} \chibar_\betadot \psi^\gamma_R  
Y_{\psibar_L}^ \alphadot }(-p,p)
\Ga_{ \psi^\beta_L   {\psibar_{L\alphadot}}}(p,-p) \nonumber \\
&- &
 \Ga_{\epsilonbar ^ {\betadot}\chibar_\betadot \psi^\beta_L  
Y_{\psi_L\alpha } }(p,-p)
\Ga_{\psi^\gamma_R   {\psi_{L}^\alpha}}(-p,p)
+ \Ga_{\epsilonbar ^ {\betadot} \chibar_\betadot \psi^\gamma_R  
Y_{\psi_R \alpha }}(-p,p)
\Ga_{ \psi^\beta_L   {\psi_{R}^ \alpha}}(p,-p)\ . \nonumber
\end{eqnarray}
Here all  vertex functions are evaluated with constant gauge coupling.
In particular the expression on the left-hand-side is nothing but the
scalar part of the ordinary 
electron self energy, where the derivative can be evaluated using the
topological formula:
\begin{equation}
 e \partial_e \Ga^ {(l)} _{\psi_L^ \beta \psi_R^ \gamma} (p,-p) = 
2l\, \Ga^ {(l)} _{\psi_L^ \beta \psi_R^ \gamma} (p,-p)  \ .
\end{equation}


Now we argue that the r.h.s.\ of (\ref{Gapsi}) can be written as
 \begin{equation}
\label{NRTpsi}
 e^ 3 \partial_e \Ga _{\psi_L^ \beta \psi_R^ \gamma} (p,-p) =
2 \Sigma_{ \psi,\rm div}(p^2)  \Ga _{\psi_L^ \beta \psi_R^ \gamma} (p,-p)
+ \epsilon_{\beta \gamma} m p^2 \Sigma_{\psi, \rm conv} \ ,
\end{equation}
where $\Sigma_{\psi, \rm conv}$ is superficially convergent and
$\Sigma_{ \psi,\rm div}$ contains the superficial divergence.
Indeed, the expressions
 $\Ga_{\epsilonbar ^ {\betadot}\chibar_\betadot \psi^\beta_L  
Y_{\psi_L\alpha } }(p,-p) $ (and similar for $\psi_R$) 
in the last line of (\ref{Gapsi}) are logarithmically divergent and
are of the same loop order as the left-hand-side.
Due to their Lorentz structure we write
\begin{equation}
\label{spsi}
 \Ga_{\epsilonbar ^ {\betadot}\chibar_\betadot \psi^\beta_L  
Y_{\psi_L\alpha } }(p,-p) = \delta^ \alpha _ \beta  \Sigma_{\psi,
\rm div} (p^2).
\end{equation}

On the other hand, the Green functions
$ \Ga_{\epsilonbar ^ {\betadot}\psi^\beta_L   Y_{\phibar_R}}(p,-p)$
and $ \Ga_{ \epsilonbar ^ {\betadot}\chibar_\betadot
\psi^\beta_L 
Y_{\psibar_R}^ \alphadot }(p,-p)  $ have dimension zero, but due to their
Lorentz structure they are superficially convergent with
the degree of divergence being $-1$. Inspection of the diagrams shows that
they vanish in 1-loop order and contribute only from  2-loop order
onwards.
In (\ref{Gapsi}) these  superficially convergent Green functions are multiplied with
linearly divergent Green functions. In  the $l$-loop
expression the divergent Green functions appear therefore
  at most with  loop  order  $l-2$. 
 Products of superficially convergent functions
 with lower order divergent Green functions contribute to the
left-hand-side as superficially convergent contributions and the
  divergences of lower orders  are
 related to divergent
 subdiagrams. Using the Lorentz structure we denote the contributions
of the first two lines with
$m \epsilon_{\beta \gamma}p^2 \Sigma_{\psi, \rm conv}$, establishing (\ref{NRTpsi}).

Explicitly, in 1-loop order (\ref{NRTpsi}) reduces to
\begin{equation}
\label{NRTpsi1loop}
  e^ 2  \Ga^ {(1)} _{\psi_L^ \beta \psi_R^ \gamma} (p,-p) =
 \epsilon_{\beta\gamma} m \Sigma^ {(1)}_{ \psi, \rm div }(p^2)    \ .
\end{equation}

For working  out the non-renormalization of chiral vertex functions
we have to show that the  function
$\Sigma_{\psi,\rm  div}$ is the only divergent contribution to 
 the chiral vertex function $\Ga_{q\psi_L
\psi_R}$.
Indeed, differentiating the
  identity (\ref{STchibar})  with respect to $ \psi_L, \psi_R$ and
$q$ and repeating the same steps as above, we obtain the identity 
\begin{eqnarray}
\label{NRTqpsi}
& &  e^ 3 \partial_e \Ga _{q \psi_L^ \beta \psi_R^ \gamma} (p_1,p_2,p_3)
= \nonumber \\
& & ( \Sigma_{ \psi, \rm div} (p_2^ 2) + \Sigma_{\psi, \rm div}(p_3^ 2))
 \Ga _{q \psi_L^ \beta \psi_R^ \gamma} (p_1, p_2, p_3) + 
\epsilon_{\beta\gamma}\Sigma_{q\psi, \rm conv}\ .
\end{eqnarray}
While the convergent part $\Sigma_{q\psi, \rm conv}$ appearing here is
different from $\Sigma_{\psi, \rm conv}$, the divergent part in
(\ref{NRTqpsi}) and (\ref{NRTpsi}) is the same.

In one-loop order one has for example
\begin{equation}
\label{NRTqpsi1}
\epsilon_{\beta\gamma}\Sigma_{q\psi, \rm conv} =
- m \Bigl( \Ga^ {(1)}_{
\epsilonbar ^ {\betadot}  \chibar_\betadot q \psi^\beta_L  
Y_{\psi_L}^ \gamma }(p_1,p_2, p_3)
- \Ga^ {(1)}_{
\epsilonbar ^ {\betadot}  \chibar_\betadot q \psi^\gamma_R  
Y_{\psi_R}^ \beta }(p_1,p_3,p_2) \Bigr)
\end{equation}
and therefore
\begin{eqnarray}
2e^2\Ga _{q \psi_L^ \beta \psi_R^ \gamma}^{(1)} (p_1,p_2,p_3)
& = & ( \Sigma^ {(1)}_{ \psi, \rm div} (p_2^ 2) + \Sigma^ {(1)}_{\psi,
\rm div}(p_3^ 2)) 
\epsilon_{\beta\gamma}
\\& -& 
 m \Bigl( \Ga^ {(1)}_{
\epsilonbar ^ {\betadot}  \chibar_\betadot q \psi^\beta_L  
Y_{\psi_L}^ \gamma }(p_1,p_2, p_3)
- \Ga^ {(1)}_{
\epsilonbar ^ {\betadot}  \chibar_\betadot q \psi^\gamma_R  
Y_{\psi_R}^ \beta }(p_1,p_3,p_2) \Bigr) \nonumber 
\end{eqnarray}
and similar expressions  in higher orders.
Therefore the only
divergent contribution to the vertex function
$\Ga_{q\psi\psi}$ is   again $\Sigma_{\psi,\rm  div}(p^2)$ 
appearing in the chiral self energy of the electrons.
 For this reason the divergences of $\Ga_{\psi \psi} $ and 
$\Ga_{q\psi\psi}$ 
can be absorbed into  the gauge-dependent part  $z_{\psi}f_\psi(\xi)$
of the electron field redefinition
(\ref{Gactinv}). Equivalently, and as expressed by eq.\ (\ref{ctidentification}), the
renormalization constants for the mass parameter $m$ and the field $q$
have to be equal.

\subsection{The non-renormalization of the photon self energy in
 higher orders}

Since divergent counterterms to the photon self energy are absent in loop order
$l \geq 2$, the photon self energy can be completely related to
non-local, counterterm independent expressions. For working out this
relation
 we will apply a similar procedure as in the previous
 section.   The result differs from the corresponding expression for
 chiral vertex function  in a remarkable way:
 The photon self energy is not 
related to superficially convergent Green functions but
to expressions which are linearly  divergent by power counting.
By gauge invariance, however, the local divergent part of all these
expressions is completely determined 
by the non-local part.

For the derivation we need two relations of SQED Green functions
determined in a previous paper \cite{HKS99}.
The first relation is the relation between 
the photon and  photino self energy
\begin{eqnarray}
\label{Sphotino}
 \Ga_{\lambdabar^ \betadot \lambda^ \alpha} (-p,p) = - \sigma^ \mu _{\alpha
 \betadot}p_\mu (1 + \Sigma_A (p^2 ))\ ,
\end{eqnarray}
where $\Sigma_A(p^2)$  is the self energy of the photon:
\begin{equation}
\label{Sphoton}
\Ga_{A^ \mu A^ \nu} (-p,p)
= -(\eta_ {\mu \nu} p^ 2 - p_\mu p_ \nu)(1+ \Sigma _A (p^2)) \ .
\end{equation}
(We use the $B$-gauge, where the photon self energy is transversal.)
Second we use that  the supersymmetry transformation of the photino into the
photon is local and determined by its tree expression:
\begin{eqnarray}
\label{susyphotino}
\Ga_{\epsilon^ \beta A_ \mu  Y_{\lambda \alpha}} (p,-p) =
p_\rho (\sigma ^ {\rho \mu})_\beta{ }^ \alpha  \ ,
\qquad 
\Ga_{\epsilonbar^ \betadot A_ \mu  Y_{\lambdabar \alphadot}} (p,-p) =
- p_\rho (\sigmabar ^ {\rho \mu}) ^\alphadot{ }_\betadot\; .
\end{eqnarray}

In a first step we extract a supersymmetry transformation
 $\delta_\alpha$ from the photon self energy by
using $\frac{\delta
 S(\Gamma)}{\delta\epsilon^\beta\delta\chi_\beta}$
 (which reproduces the  complex conjugate identity to 
(\ref{STchibar})). With eq.~(\ref{eetabar})  and with the help of  
 the topological formula 
 we find the following result:
\begin{equation}
\label{Gaphoton}
2 (l-1)e^ 2 \Ga_{A^ \mu A^ \nu} ^ {(l)}  =  
p^\rho  \Ga^ {(l)}_{\chi_\beta A^ \nu  \lambda^ \alpha} (-p , p) 
(\sigma_ {\mu \rho })_{\beta }{ } ^ \alpha 
\! -  \! p^ \rho \Ga^ {(l)}_{\chi_\beta A^ \mu  \lambda^ \alpha} (p ,- p) 
(\sigma_{\nu \rho}) _{\beta }{ }^{ \alpha }  .
\end{equation}
As for the invariant counterterms (cf.~(\ref{Gactinv})),
 the one-loop level is distinguished from higher loop orders in the
  expression (\ref{Gaphoton}):
For $l =1$
 the photon self energy drops out, and it cannot be related to a
 non-local, renormalization-independent expression.
It is divergent and the divergences are absorbed by the counterterm
$z_{\rm kin} {(1)}$.
At the 1-loop level
 the identity (\ref{Gaphoton}) together with the gauge Ward identity
 instead determines the vertex function 
$\Ga^ {(1)}_{\chi_\beta A^ \nu  \lambda^ \alpha} $:
\begin{equation}
\Ga^ {(1)}_{\chi_\beta A^ \nu  \lambda^ \alpha} (-p , p) = 0.
\end{equation}

In higher orders eq.~(\ref{Gaphoton})
 relates the photon self energy 
 to the Green functions $\Ga^ {(l)}_{\chi_\beta A^ \mu  \lambda^
 \alpha}  $.
Moreover, we are able to extract a supersymmetry transformation $\bar
 \delta_\alphadot$ 
from the chiral Green function $\Ga^ {(l)}_{\chi_\beta
A^ \mu  \lambda^ 
 \alpha}  $ 
by testing the identity (\ref{STchibar2}) with respect to
$\lambda^ \beta, A^ \mu$ and $ \chi^ \gamma$.
Using the topological formula (\ref{topfor})  and the relation
\begin{equation}
\Ga_{\epsilonbar^ \betadot \chibar _\betadot  \lambda^ \beta
Y_{\lambda\alpha}} (-p,p) =
 \delta _{\beta}^ {\alpha}\ ,
\end{equation}
which follows from the linear equation (\ref{EquivNilpotency}), we find
in momentum space 
\begin{eqnarray}
\label{GachilambdaA}
2 e ^2 l\; \Ga^ {(l)}_{\chi_\gamma \lambda^ \beta A^ \mu  } (-p , p)
& = & 
i \sigma ^ \rho _{\beta \betadot} \Ga^ {(l)}_{\chibar_\betadot \chi_ \gamma
A^ \mu A ^\rho} (p,-p) - p_\rho (\sigmabar^ {\rho \mu})^ \alphadot{ }_\betadot
\Ga^ {(l)}_{\chibar_\betadot \chi_ \gamma
\lambda ^ \beta  \lambdabar ^ \alphadot} (- p, p) \nonumber \\
& & - \sum_{k= 0}^ l
\Ga^ {(l-k)}_{\epsilonbar^ \betadot
\chibar_\betadot \chi_ \gamma
A^ \mu Y_{\lambdabar}^ {\alphadot}} (p, -p) \Ga^ {(k)}_{\lambda^ \beta
\lambdabar_\alphadot } ( -p , p).
\end{eqnarray}
Used in eq.~(\ref{Gaphoton}) 
this expression determines  the photon self energy in $l \geq 2 $.

Let us discuss the right-hand-side of eq.~(\ref{GachilambdaA}).
On the one hand, the naive degree of power counting  is not
improved, but 
 the
individual parts are   linearly and logarithmically divergent. On the
other hand, however,
all  Green functions appearing with the photon field are constrained by the
gauge Ward identity. In this way
 all local divergent contributions are uniquely 
determined by
  non-local, convergent contributions, as we will now show.

The most important term in (\ref{GachilambdaA}) is the linearly divergent Green
function $\Ga_{\chibar \chi
 A^ \mu  A ^\nu} $. From the gauge Ward identity one finds
\begin{equation}
\label{transversal}
 p_1^ \mu\Ga_{\chibar_\betadot \chi_ \gamma
A^ \mu A ^\nu} (p_3, p_4 , p_1, p_2) = 
p_2^ \nu\Ga_{\chibar_\betadot \chi_ \gamma
A^ \mu A ^\nu} (p_3, p_4 , p_1, p_2) = 0 \; ,
\end{equation}
where temporarily non-zero momenta $p_3$ and $p_4$ have been assigned to the $\chi$- and
$\chibar$-vertices. 
Taking into account parity conservation
 we find the following covariant tensor decomposition for the
 symmetrized expression
\begin{eqnarray}
\label{Gachichibarphoton}
& &\frac 12  \bigl(\Ga_{\chibar^ \betadot \chi^  \gamma
A^ \mu A ^\nu}(p_3 ,0,p_1,p_2) +
 \Ga_{\chibar^ \betadot \chi^ \gamma
A^ \mu A ^\rho}(0,p_3,p_1,p_2) \bigr) \nonumber\\
&= & 
 i \sigma^ \lambda_{\gamma \betadot}
\bigl(\epsilon _{\lambda \mu \sigma  \rho} p_1 ^ \rho
(\eta^ \sigma{ } _{\nu}  \Sigma_1 (p_1,p_2)
-   p_2^ \sigma p_1^ {\nu}
 \Sigma_2 (p_1,p_2) -
p_2^ \sigma p_{2 \nu}  \Sigma_3 (p_1,p_2)) \nonumber \\
& & \phantom{ i \sigma^ \lambda}
+ \epsilon _{\lambda \nu \sigma  \rho} p_2 ^ \rho
(\eta^ \sigma{ } _{\mu}  \Sigma_1 (p_2,p_1)
-   p_1^ \sigma p_{2{\mu}}
 \Sigma_2 (p_2,p_1) -
p_1^ \sigma p_{1 \mu}  \Sigma_3 (p_2,p_1)) \nonumber \\
&  & \phantom{ i \sigma^ \lambda} + \epsilon _{\mu \nu \sigma  \rho} p_1^ \rho p_2^ \sigma 
( p_{1\lambda}\Sigma_4
 (p_1,p_2)
+ p_{2\lambda} \Sigma_4 (p_2,p_1)) \bigr) .
\end{eqnarray}
For $p_1\ne p_2$ transversality (\ref{transversal}) restricts the single parts
appearing in the above 
expression,
\begin{equation}
\label{Sigma1}
\Sigma_1 (p_1,p_2) = p_1\cdot p_2 \Sigma_2 (p_1,p_2) +  p_{2 }^ 2
\Sigma_3(p_1,p_2) \ ,
\end{equation}
but due to analyticity (\ref{Sigma1}) has to hold also for $p_1=p_2$.
The linearly divergent  part $\Sigma_1$ in $\Ga_{\chibar_\betadot \chi^ \gamma
A^ \mu A ^\nu}(0 ,0,-p,p)$ is therefore completely determined by the
non-local parts $\Sigma_{2}$ and $\Sigma_{3}$.    

In a similar way we find
\begin{equation}
\label{GachichiY}
\Ga_{\epsilonbar^ \betadot
\chibar_\betadot \chi^ \gamma
A^ \mu Y_{\lambdabar}^ {\alphadot}} (p, -p) =
i \sigma^ \rho_{\gamma \alphadot}(\eta_{\rho\mu} p^2 - p_\rho p_\mu)
\Sigma_5(p^2) \ .
\end{equation}
Again the local part is completely determined by the non-local part. 
Finally we decompose also the logarithmically divergent Green function
 $ \Ga_{\chibar_\betadot \chi_ \gamma
\lambda ^ \beta  \lambdabar ^ \alphadot} (- p, p)$:
\begin{eqnarray}
\label{Gachichilambda}
\Ga_{\chibar^\betadot \chi^ \gamma
\lambda ^ \beta  \lambdabar ^ \alphadot} (- p, p) & = &
\epsilon_{\gamma \beta} \epsilon_{\alphadot \betadot} \Sigma_6(p^2) 
\nonumber \\
&  + &
\sigma^ \lambda_{\gamma \alphadot} \sigma^ \rho_{\beta
\betadot}p_{\rho}p_{\lambda} \Sigma_7(p^2) +
\sigma^ \lambda_{\beta \alphadot} \sigma^ \rho_{\gamma
\betadot}p_{\rho}p_{\lambda} \Sigma_8(p^2).
\end{eqnarray}
Inserting (\ref{Gachichilambda})  into (\ref{GachilambdaA})
 the first term drops out, and only 
 the non-local parts $\Sigma_7(p^2) + \Sigma_8(p^2 ) $
 contribute to the photon self energy.

When we insert the above expressions (\ref{Gachichibarphoton}) with
(\ref{Sigma1}),
(\ref{GachichiY}) and (\ref{Gachichilambda}) into
eq.~(\ref{GachilambdaA}), we find a non-local regularization
independent expression for the vertex function 
$\Ga_{\chi \lambda A}$. When  we finally  use the relation (\ref{Gaphoton})
we find the desired result, in which  the 
photon self energy for $l\geq 2$ is completely expressed by non-local
Green functions: 
\begin{eqnarray}
l(l-1) e^4 \Sigma^ {(l)}_A (p^2)
& = & p^2
\bigl(-4( \Sigma^ {(l)}_2(p^2) -  \Sigma^ {(l)}_3 (p^2)) + \Sigma^
{(l)}_7 (p^2) +  
\Sigma^ {(l)}_8 (p^2)  \\
& & 
 + \Sigma_5^ {(l)} + \sum_{k= 1}^ {l-2}\Sigma^ {(l-k)}_5(p^2)  \Sigma^
 {(k)}_A(p^2)\bigr) \nonumber 
\end{eqnarray}
Here we have used the supersymmetry relation between photon and
photino self-energy (\ref{Sphotino}), which makes possible to
express  the self energy of the
photino appearing in (\ref{GachilambdaA})
by  the photon self energy.
From 3-loop order onwards the divergent 1-loop self energy of the
photon
appears in
the above formula, but only in a product with the non-local
contribution $\Sigma_5$. Such an expression does not contribute to
superficial divergences of the photon self energy in 
$l\geq 2 $ but is related to the appearance of 
subdivergences.
 

The arguments we have used here for relating the photon self energy to
non-local expressions are
  analogous to the ones used for calculating 
the axial-current--photon--photon Green functions (see \cite{AD69} and
 \cite{JE00} for a recent review). There, too, the divergent part of
 the triangle diagram is uniquely related to non-local expressions by
 gauge invariance. In fact, there is a connection
 between the photon self energy and the
axial-vector Green functions and the Adler--Bardeen anomaly. This is
revealed when the SQED action with local coupling is extended by the axial
vector multiplet. Then we will see that the $\chibar \chi$-insertions
contributing on the right-hand-side  of (\ref{GachilambdaA}) induce
nothing but 
insertions of the axial current (see section 9).

\section{Softly broken axial symmetry and the axial vector multiplet}
\label{SecAxial}

Although the perturbative construction of SQED with local couplings is
well-defined, it is not complete in the sense of multiplicative
renormalization.
 The invariant counterterm $\Ga_{\rm ct, matter}$ (\ref{ctmatter})
 cannot be
related to  a wave function renormalization as long as the coupling
is local.  Therefore it is impossible to interpret all free parameters
of higher orders as field and coupling renormalizations. In this
section 
multiplicative renormalizability is restored by introducing an
additional axial vector multiplet. Then the counterterm  $\Ga_{\rm ct,
  matter}$ (\ref{ctmatter}) can 
be understood as a field renormalization of the axial vector multiplet
and the matter fields, and it
is possible to derive the Callan--Symanzik equation.

The axial symmetry 
\equa{
\label{matteraxial}
 \delta^ 5 \phi_A & = -i \tilde \omega(x) \phi_A \ ,
\qquad \delta^ 5 {\bar \phi}_A  = i \tilde \omega(x) \bar \phi_A\ ,
\qquad \phi = \varphi, \psi, F; \quad A = L,R \nonumber
 \\
\delta^ 5 A^\mu & = \delta^ 5 \lambda  = \delta^ 5 \lambdabar  =
\delta^ 5 D = 0
}
is softly broken by the matter mass term  and gives rise to a
partially conserved axial current in the classical approximation:
\begin{equation}
\delta^ 5 \Ga_{\rm SQED} = - \intd  \tilde \omega (x) \partial^ \mu  
j_ \mu^ {\rm axial} - 2 i m\intd \tilde \omega(x) (L_{\rm mass} - \bar L_{\rm mass})
\label{jaxial}
\end{equation}
with the axial current
\begin{eqnarray}
j_{\mu} ^ { \rm axial} = & i \bigl( \phibar _L D_ \mu
 \varphi_L - \varphi_L D_\mu \phibar _L    + i
\psi_L  \sigma_\mu \psibar_L + (_{L\to R}) \bigr) \ . 
\end{eqnarray}
(We take the local coupling in all covariant derivatives.)

Although the axial symmetry is softly broken, it can be  gauged and
supersymmetrized in the same way as electric charge symmetry:
We introduce the external axial multiplet $V^ \mu, \lambdaV^ \alpha,
\lambdaVbar^ \alphadot, \DV$
with axial transformations
\begin{equation}
\label{Vaxial}
\delta^ 5 V^ \mu =  \partial ^ \mu \tilde \omega (x) ,
\qquad
\delta^ 5 \lambdaV^ \alpha = \delta^ 5 \lambdaVbar^ \alpha = 
\delta^ 5 \DV = 0,
\end{equation}
and supersymmetry transformations in the Wess--Zumino gauge
\equa{
\delta_\alpha V^ \mu & = i \sigma_{\alpha \alphadot} ^ \mu \lambdaVbar 
^ {\alphadot}, &\quad &\bar \delta_{\alphadot}
V^ \mu  = - i \lambdaV  
^ {\alpha}\sigma_{\alpha \alphadot} ^ \mu, \nonumber \\ 
\delta_\alpha \lambdaV^ { \beta} & = 
\frac i 2 \sigma^{\mn \; \beta}_\alpha  F_{\mn}(V)
+ \frac i 2 \delta_\alpha^ \beta \DV,  &\quad &\bar \delta_{\alphadot} \lambdaV
_{\beta} = 0 ,\nonumber \\
\delta_\alpha \lambdaVbar_\alphadot & = 0 , &\quad &\bar\delta^ {\alphadot}
\lambdaVbar _{\dot \beta}=\frac i 2 {\bar \sigma}^ {\mn \; \alphadot }_
{\phantom{\mn; \alpha} \dot \beta}
  F^ {\mn} (V)
+ \frac i 2 \delta^ \alphadot_ {\dot\beta} \DV, \nonumber \\
\delta_\alpha \DV & = 
2 \sigma^ \mu_{\alpha \alphadot} \partial _\mu \lambdaVbar ^{ 
\alphadot} , &\quad &\bar \delta_{\alphadot} \DV = 
2 \partial _\mu \lambdaV ^ { 
\alpha}\sigma^ \mu_{\alpha \alphadot} .
}
When we extend the covariant derivatives to
\begin{equation}
\label{covDV}
D^ \mu \phi_A = (\partial ^ \mu + i e(x) Q_A A^ \mu + i V^ \mu)\phi_A\
, \qquad A = L, R 
\end{equation}
in the matter part of the action
\begin{eqnarray}
\Ga_{\rm matter} 
& = &\intd \Bigl(
D^ \mu  \phibar_L D_\mu \varphi_L+ i   \psi_L^ \alpha \sigma^ \mu
_{\alpha \alphadot}  D_\mu \psibar_L
^ {\alphadot}
+ {\overline F}_L  F_L \nonumber \\
& & \;+ i e Q_L  \sqrt{2}( \lambda \psi_L  \phibar_L  -  \lambdabar  \psibar_L
 \varphi_L) 
+ i  \sqrt{2}( \lambdaV \psi_L  \phibar_L  -  
\lambdaVbar  \psibar_L
 \varphi_L) \nonumber \\ 
& & \; + \frac 1 2 e Q_L D \phibar_L \varphi _L + \frac 1 2 \DV
\phibar_L \varphi_L + (_{L \to R}) \Bigr)
\label{GamatterV}
\end{eqnarray}
and in the supersymmetry transformations, then the kinetic part
(\ref{Gakinext}) and the matter part (\ref{GamatterV})
are invariant under gauge symmetry with local coupling (\ref{gaugetr}), axial
symmetry (\ref{matteraxial}, \ref{Vaxial}) and supersymmetry
transformations. 

Finally we assign shifted axial transformations
 to  the neutral chiral  multiplet ${\bf q} = (q, q^ \alpha, q_F)$
 (\ref{qdef}) and its complex conjugate:
\begin{equation}
\label{qaxial}
\delta^ 5 {\mathbf q} = 2 i ({\mathbf q} + (m,0,0))  \tilde \omega
 (x)\ ,
\qquad \delta^ 5 {\mathbf {\bar q}} = - 2 i ({\mathbf {\bar q}}  +
 (m,0,0)) \tilde \omega(x)
\end{equation}
and replace the ordinary derivative by
\begin{equation}
\label{covq}
D_\mu {\mathbf q} = (\partial _\mu - 2 i V_ \mu)({\mathbf q} + (m,0,0))
\end{equation}
in the supersymmetry transformations of the ${\mathbf q}$
multiplet. Then the sum of the matter  mass term $\Ga_{\rm mass}$ (\ref{Lmass})
and the chiral ${\mathbf q}$-interaction term (\ref{Gaq}) become 
invariant under axial symmetry
\begin{equation}
\delta^ 5 (\Ga_{\rm mass} + \Ga_q) = 0,
\end{equation}
and axial symmetry can be used as a defining symmetry of the model.

For quantization the axial symmetry (\ref{WIaxial}) has to
 be included into the
Slavnov--Taylor 
identity in the same way as the gauge transformations (see section 3.1).
The respective BRS transformations are obtained
by replacing the transformation parameter $\tilde \omega (x)$ with the
 parity odd axial ghost  $\tilde c (x)$.
The  BRS transformation of the axial ghost is 
determined by the structure constants of the algebra:
\begin{equation}
s \tilde c = 2 i \epsilon \sigma \epsilonbar V - i \omega \partial
\tilde c .
\end{equation}
The complete BRS transformations of SQED with local coupling and the
axial vector multiplet
 are summarized in  appendix C.

The Slavnov--Taylor operator of the extended
action includes 
 the BRS transformations of the axial vector multiplet 
$(V^ i = (V^ \mu, \lambdaV^ \alpha,
\lambdaVbar^ \alphadot, \DV))$ and of the axial
 ghost 
in addition to (\ref{STOperator}): 
\begin{equation}
\label{STOperatoraxial}
{\cal S} ({\cal F}) \to
{\cal S} ({\cal F})  + \intd \Bigl(
 s V^ i \dF{V^ i} + s\tilde {c} \dF{\tilde {c}} \Bigr).
\end{equation}
Nilpotency of the Slavnov--Taylor operator (\ref{nilpotency}) and the linear equation
(\ref{EquivNilpotency})
remain valid in their original form.
The classical action with gauged axial symmetry
satisfies the Slavnov--Taylor identity with the extended
Slavnov--Taylor operator
(\ref{STOperatoraxial}) 
by construction: 
\begin{equation}
\label{STidentityaxial}
{\cal S} (\Gacl) = 0\ .
\end{equation}
Similarly as in eq.\ (\ref{ghostequ}) for the Faddeev--Popov ghost $c(x)$,  
a linear ghost equation for the axial ghost can be derived:
\begin{equation}
\label{axialghost}
\frac {\delta \Gamma} {\delta {\tilde c}}  = 
i  (  \psi_ {L}^ {\alpha} Y_{\psi_L  \alpha} +
\varphi_L Y_{\varphi_L} - \psibar_{L \alphadot} Y_{\psibar_L}^ {
\alphadot} -
\phibar_L Y_{\phibar_L}) + ({}_{L\to R})  \ .
\end{equation}
Using the consistency equation 
\begin{equation}
\frac \delta {\delta \tilde c}
{\cal S}(\Ga) +  s_{\Ga} 
\bigl(\frac \delta {\delta \tilde c} \Ga \bigr) =  \partial^\mu  
\frac {\delta} {\delta {V^\mu}}\Ga
\label{axialghostcons}
\end{equation}
at the classical level together with (\ref{STidentityaxial}) we can
derive an axial Ward identity expressing softly broken axial symmetry:
\begin{eqnarray}
\Bigl( {\bf w}^ {\rm axial} - \partial^\mu  
\frac {\delta} {\delta {V^\mu}}\Bigr)\Ga_{\rm cl} &  = & - 2 i m
\Bigl( \frac 
{\delta \Ga_{\rm cl}} {\delta q} - \frac {\delta \Ga_{\rm cl}} {\delta
 \qbar} \Bigr) 
 +{\cal O}(\omega) 
\label{WIaxial}
\end{eqnarray}
with 
\begin{eqnarray}
{\bf w}^{\rm axial} & = & -i\Bigl(\varphi_L\dfunc{}{\varphi_L} -
Y_{\varphi_L}\dfunc{}{Y_{\varphi_L}} + \psi^\alpha_L\dfunc{}{\psi^\alpha_L}
- Y^\alpha_{\psi_L}\dfunc{}{Y^\alpha_{\psi_L}} 
\nonumber\\&&{}
- \phibar_L\dfunc{}{\phibar_L}
 + Y_{\phibar_L}\dfunc{}{Y_{\phibar_L}}
 - \psibar^\alphadot_L\dfunc{}{\psibar^\alphadot_L}
 + Y^\alphadot_{\psibar_L}\dfunc{}{Y^\alphadot_{\psibar_L}} \Bigr) + (_{L\to R}) \nonumber \\&&{}
+ 2 i \Bigl( q \dfunc{}{q}
+  q^ \alpha\dfunc{}{q^ \alpha}+  q_F \dfunc{}{q_F}
 -  {\qbar }\dfunc{}{{\qbar}}-  {\qbar^ \alphadot}\dfunc{}{{\qbar^
\alphadot }} 
- {\qbar_F}\dfunc{}{{\qbar_F}}  \Bigr)\ .
\end{eqnarray}

\section{The Adler--Bardeen anomaly and \\ its non-renormalization theorem}

At the quantum level the axial Ward identity (\ref{WIaxial}) is broken  by the
Adler--Bardeen anomaly \cite{AD69, BA69, BEJA69}. Since  the axial
transformations are included  
in the Slavnov--Taylor identity,
 the anomaly will  appear  as an
anomaly of the Slavnov--Taylor identity in the present construction.
In one loop order, all breakings of the Slavnov--Taylor identity
\begin{equation}
{\cal S}(\Ga) = \Delta + {\cal O}(\hbar ^2)
\end{equation}
 are
algebraically  restricted by
 the nilpotency
properties of the Slavnov--Taylor operator (\ref{nilpotency})
and by the identity (\ref{holomorph}),
which restricts  the  appearance of $\eta-\etabar$ to a total derivative.
Therefore $\Delta$ has to satisfy the following constraints:
\begin{equation}
\label{Sdelta}
s_{\Ga_{cl}} \Delta = 0 \qquad \mbox{and} \qquad
\intd \Bigl(\frac {\delta } {\delta \eta}  - \frac {\delta} {\delta \etabar}
\Bigr) \Delta = 0\ .
\end{equation}
Moreover, $\Delta$ is restricted by  discrete symmetries (see appendix
C) and the
number of local couplings appearing in the field monomials are
determined by the 
\footnote{
The topological formula (\ref{topfor}) is valid
also in presence of the axial vector multiplet.}.

The anomaly is a field polynomial that appears in $\Delta$ but cannot
be absorbed by a counterterm contribution. 
Setting the supersymmetry ghosts to zero,
 the anomaly appears with the ghost of axial symmetry and has the
following form:
\begin{equation}
\label{gaugeanomalie}
\Delta^ {\rm anomaly}\Big|_{\epsilon, \epsilonbar, \omega= 0}  =  
r^ {(1)}\intd\tilde c  \epsilon^{\mu \nu \rho\sigma}
F_{\mn} (eA )
F_{\rho\sigma} (eA) \ .
\end{equation}
Using the consistency equation (\ref{axialghostcons}) 
we find that the same anomaly appears in
 the axial Ward identity (\ref{WIaxial}) and that $r ^ {(1)}$ is determined
 by the usual triangle diagram:
\begin{equation} 
\label{r1}
r^ {(1)} = - \frac {1} {16 \pi^ 2} \ .
\end{equation}

It is a crucial feature of the construction with the local coupling
and axial vector field, that
the axial anomaly (\ref{gaugeanomalie}) can be written as a
differential operator acting on the classical action:
\begin{equation}
\label{susygaugeanomaly}
\intd \tilde c \bigl ( \epsilon^{\mu \nu \rho\sigma}
F_{\mn} (eA )
F_{\rho\sigma} (eA) - 4 \partial _ \mu(\lambda \sigma^ \mu
\lambdabar)\bigr) =   
4 i  \intd  \Bigl(\tilde c \Bigl(\frac {\delta}{\delta \eta} -
\frac {\delta}{\delta \etabar}\Bigr) \Gacl \ .
\end{equation}

Supersymmetry  determines a supersymmetric extension of the
Adler--Bardeen
anomaly \cite{White92a, MPW96a}. 
In our framework, its form can be found 
 by extending the right-hand-side of (\ref{susygaugeanomaly}) 
to a BRS-invariant
operator. As a result, the Adler--Bardeen anomaly and its 
supersymmetric extension   can
be written as a BRS-invariant operator acting on the classical action:
\begin{eqnarray}
\label{operatoranomalie}
\Delta ^ {\rm anomaly}
& = & 4 i r^{(1)} \intd  \Bigl(\tilde c \Bigl(\frac {\delta}{\delta \eta} -
\frac {\delta}{\delta \etabar}\Bigr) + 2i (\epsilon \sigma^ \mu )^
\alphadot V_\mu \frac {\delta}{\delta \chibar^ \alphadot} 
 - 2i ( \sigma^ \mu \epsilonbar )^ \alpha V_\mu \frac
 {\delta}{\delta \chi^ \alpha} \nonumber \\ { } & &
 + 2 \epsilonbar_\alphadot \lambdaVbar ^\alphadot 
\frac {\delta}{\delta f} - 2 \lambdaV^\alpha \epsilon_\alpha \frac {\delta}{\delta
 \fbar} \Bigr) \Ga_{\rm cl} \nonumber \\
& \equiv & - r^ {(1)} \delta {\cal S}\Ga _{\rm cl}\ .
\end{eqnarray}

Since the anomaly $\Delta^{\rm anomaly}$
can be expressed in this form, it can be absorbed into a redefined
Slavnov--Taylor operator:
\begin{equation}
\label{SToperatoranom}
 {\cal  S} (\Ga) + r^{(1)} \delta
{\cal S} \Ga =  0 + {\cal
O}(\hbar^ 2)  \ .
\end{equation}
The new piece $\delta {\cal S}$ being a symmetric operator,
the operator $  \bigl({\cal S}+ r^  {(1)}\delta {\cal S} \bigr) (\cal F)$ 
and its linearized
version $ s_{\cal F}  + r ^ {(1)}\delta {\cal S}$ have the 
same nilpotency properties as the original Slavnov--Taylor operator
(see (\ref{nilpotency})). Owing to this property, 
algebraic renormalization can be continued to higher orders in spite
of the presence of the anomaly.

In particular,
the breakings of higher orders are restricted by (\ref{Sdelta})
in the same way as
 the ones of 1-loop order.
Taking into account  
the topological formula,  the gauge anomaly of loop order  $l$
takes the  general form:
\begin{equation}
\Delta_l^ {\rm anomaly}\Big|_{\epsilon, \epsilonbar, \omega= 0}  = 
r^ {(l)} \intd e^ {2(l-1)}(x)\tilde c  \,  \epsilon^{\mu \nu \rho\sigma}
F_{\mn} (eA )
F_{\rho\sigma} (eA) \ .
\label{anomalygeneral}
\end{equation}
Apparently the 1-loop case $l=1$  is special.
At this order the coefficient of the ghost in (\ref{anomalygeneral})
is a total derivative, but with local coupling it is not a total
derivative at higher orders. Hence, when we use
 the consistency equation 
(\ref{axialghostcons}) and integrate the axial Ward identity,
the anomaly appears in the
 the  Ward identity of global axial symmetry for $l\geq 2$
\begin{equation}
\label{axWIanomaly}
{\cal W}^ {\rm axial} \Ga = r^ {(l)} \intd
e^ {2(l-1)}(x)
\epsilon^{\mu \nu \rho\sigma}
F_ {\mn} (eA )
 F_ {\rho \sigma} (eA)  + {\rm soft\ terms} 
\end{equation}
with 
\begin{equation}
{\cal W}^ {\rm axial} \equiv \intd {\mathbf w}^ {\rm axial}(x) \ .
\end{equation}
Only for $l=1$ the integrand in (\ref{axWIanomaly}) is a total
derivative and the integral vanishes.

Owing to the structure of $ {\cal W}^ {\rm axial} $, no term on the
l.h.s.\ of (\ref{axWIanomaly}) is composed solely of photon fields. 
Hence, testing with respect to two photon fields shows that
 the coefficients $r^ {(l)}$ have to
vanish
identically for $l\geq 2$:
\begin{equation}
r^ {(l)} = 0 \quad \mbox{ if } \quad l\geq 2.
\end{equation}
Local couplings provide therefore a very simple and elegant way for
proving 
 the non-renormalization of the
axial
anomaly \cite{ADBA69}. It is clear that the  derivation is not
restricted to supersymmetric gauge theories but can be applied in the same way to
ordinary gauge theories.

In summary we find that
the Green functions of SQED with local gauge coupling and an axial
vector multiplet satisfy  the anomalous Slavnov--Taylor identity 
\begin{equation}
\label{STanomalous}
\bigl( {\cal  S}  + r^{(1)} \delta
{\cal S}\bigr) \Ga =  0 
\end{equation} 
and the anomalous axial Ward identity 
(cf.~(\ref{WIaxial})):
\begin{eqnarray}
\Bigr( {\bf w}^ {\rm axial} - \partial^ \nu \frac {\delta}{\delta V ^\nu} 
\Bigl) \Ga &  = & 4 i r^ {(1)}
 \Bigl(\frac {\delta } {\delta \eta}  - \frac {\delta} {\delta \etabar}
\Bigr) \Ga  
 - 2 i m
\Bigl( \frac 
{\delta } {\delta q} - \frac {\delta } {\delta \qbar}
\Bigr)\Ga +
{\cal O}(\omega)
\label{WIaxialanomalous}
\end{eqnarray}
to all orders of perturbation theory. 

We want to note already here that
the supersymmetric extension of the Adler--Bardeen anomaly induces a
modification of  
 supersymmetry  transformations for axial-current
Green functions. A detailed discussion
of this point and its relation to the manifestly  supersymmetric gauge
can be found in  section 9.

\section{The Callan--Symanzik equation}

In this section the power of the construction using the local gauge coupling
and the anomalous  axial current becomes evident. 
With the axial multiplet all invariant counterterms can
be given as invariant field operators acting on the classical action;
consequently the model possesses a valid Callan--Symanzik equation.
 The non-renormalization theorems derived in section 3 and 4 will
 then appear as
restrictions on the various Callan--Symanzik coefficients. 
Indeed, we
will show that the Callan--Symanzik equation 
 contains only two gauge-independent, physical coefficients:
the 1-loop $\beta$-function of the gauge coupling
 and the anomalous dimension of the mass, $\gamma$.
In particular the construction provides  the 
$\beta$-function of the gauge coupling in its closed form.

\subsection{The CS equation with local coupling}

We start the construction of the Callan--Symanzik (CS) equation with
 local
 couplings in
the classical approximation:
\begin{eqnarray}
\label{CScl}
 \mu_i\partial_{ \mu_i} \Gacl & = &
- 2 m^2 (\varphi_L \phibar _L + \varphi_R \phibar _R) -m (\psi_L
\psi_R+ \psibar_L \psibar_ R) \\
& = & m \intd \Bigl(\frac {\delta} {\delta q} + \frac {\delta} {\delta
\qbar}\Bigr) \Gacl \ .\nonumber
\end{eqnarray}
The mass differentiation $  \mu_i\partial_{ \mu_i}$ contains the
differentiation with respect to all mass parameters of the theory
($\kappa$ is a normalization point introduced for an off-shell
normalization of the matter field residua):
\begin{equation}
\mu_i\partial_{\mu_i} = m\partial_m + \kappa \partial_\kappa \ .
\end{equation}
Eq.~(\ref{CScl}) can be rewritten as
\begin{equation}
\label{CScl2}
\mu_i D _{\mu_i} \Gacl = 0
\end{equation}
with the operator
\begin{equation}
\mu_i D _{\mu_i} \equiv
m \partial_m + \kappa \partial_\kappa - m
\intd \Bigl(\frac {\delta} {\delta q} + \frac {\delta} {\delta
\qbar}\Bigr)\ ,
\end{equation}
which is  symmetric with respect to the anomalous Slavnov--Taylor identity
(\ref{STanomalous}):
\begin{equation}
( s_{\Ga} + r ^{(1)} \delta{\cal S} ) \mu_i D_{\mu_i} \Ga
 = \mu_i D _{\mu_i} ( {\cal S} + r ^{(1)} \delta
 {\cal S})(\Ga) = 0  \ .
\label{CSclST}
\end{equation}

Let us briefly outline the construction of the CS equation in higher
orders before entering the calculational details.
In higher orders eq.~(\ref{CScl2}) is broken by hard terms, the
dilatational anomalies:
\begin{equation}
\mu_i D _{\mu_i} \Ga =
  \Delta_m \cdot \Gamma \ .
\label{CSbreak}
\end{equation}
Since the operator $\mu_i D_{\mu_i}$ is symmetric with respect to the
symmetries of $\Ga$,
the dilatational anomalies 
are completely characterized by the symmetries of $\Ga$:
\begin{equation}
\bigl( s_\Ga + r^ {(1)} \delta{\cal S} \bigr) \bigl(  \Delta_m \cdot \Gamma
\bigr) = 0 \, , \qquad 
\intd \Bigl(\frac {\delta } {\delta \eta}  - \frac {\delta} {\delta \etabar}
\Bigr) \bigl( \Delta_m \cdot \Gamma \bigr)  =  0 \ .  
\end{equation}
In the Callan-Symanzik equation  the breakings $\Delta_m \cdot \Ga $
are absorbed by 
symmetric  operators acting on $\Ga$. These operators can be combined
to the Callan-Symanzik
operator ${\cal C} =  \mu_i D _{\mu_i} + {\cal O}(\hbar)$, which
possesses the same symmetry properties as $\mu_i D_{\mu_i}$, in
particular it is an $s_\Ga + r^ {(1)} \delta{\cal S} $-symmetric
operator:
\begin{equation}
\label{CSSTcons}
\bigl( s_{\Ga} + r ^{(1)} \delta {\cal S}\bigr) {\cal C }   \Ga = 
{\cal C}\bigl({\cal S}+ r^ {(1)}\delta {\cal S}\bigr) \Ga +
\bigl(s_{\Ga} + r ^{(1)} \delta {\cal S}\bigr) \Delta_Y \ .
\end{equation}
The expression $\Delta_Y $  is defined to be a collection of field monomials which are
 linear in propagating fields. As such, $\Delta_Y$ appears  as a
trivial insertion  and does not need to  be absorbed into an
operator. 

The construction of the Callan-Symanzik equation proceeds by
induction. 
When the breaking $\Delta_m$
is absorbed into a symmetric Callan-Symanzik operator to loop order
$l-1$, the breaking of loop order $l$ is a  local field
monomial:
\begin{equation}
{\cal C}^{(l-1)} \Ga = 0 + \Delta_Y + {\cal O} (\hbar^ {l}) \quad
\Rightarrow \quad 
{\cal C}^{(l-1)} \Ga = \Delta_{m}^ {(l)} + \Delta_Y + {\cal O}(\hbar ^
{l+1}) \ .
\end{equation}
The local field monomial $\Delta_m^ {(l)}$ is characterized 
by the symmetries of the classical
action:
\begin{eqnarray}
s_{\Ga_{\rm cl}} \Delta_m^ {(l)}  =  0 \qquad \mbox{and } 
\intd \Bigl(\frac {\delta } {\delta \eta}  - \frac {\delta} {\delta \etabar}
\Bigr) \Delta_m^ {(1)}  =  0 \ ,
\end{eqnarray}
and the numbers of local couplings in $\Delta^ {(l)}_m$ are constrained by the
topological formula (\ref{topfor}).

Since $\Delta_m^ {(l)}$ has the same quantum numbers and the same
symmetries
as the  invariant 
counterterms of loop order $l$ (see (\ref{ctconstraints})),
it consists of the same type of field
monomials as  the invariant counterterms
(\ref{Gactinv})\footnote{The 
axial multiplet does not give rise to new counterterms except for
contributions depending solely on  external fields and are entirely
 local. They  are not
 relevant in the present context and are omitted from all equations.}:
\begin{equation}
\label{Deltambasis}
\Delta_m^ {(l)} \in
\bigl[
 \Ga^ {(1)}_{\rm ct, kin} \delta_{l1},
\Ga^ {(l)}_{\rm ct, matter}, 
\Ga^ {(l)}_{{\rm ct}, \varphi}(\xi) ,
\Ga^ {(l)}_{{\rm ct}, \psi}(\xi),
\Ga^ {(l)}_{{\rm ct}, \psi\varphi} (\xi) \bigr]\ .
\end{equation}
Suppose, we can express the invariant counterterms in
(\ref{Deltambasis}) as
$s_\Ga + r^ {(1)} \delta {\cal S}$-symmetric operators in the sense of
eq.\ (\ref{CSSTcons}). Then $\Delta_m^{(l)}$ can be absorbed by an
additional piece of loop order $l$ in the CS operator without
destroying (\ref{CSSTcons}). This establishes the CS equation
\begin{equation}
{\cal C} \Ga = \Delta_Y
\end{equation}
 by
induction. 

Hence, the derivation of the CS equation and the non-renormalization
theorems is reduced to the algebraic problem of extending the
counterterms (\ref{Deltambasis}) to  $s_\Ga + r^ {(1)} \delta {\cal
  S}$-symmetric operators.

\subsection{Construction}

For extending  the
counterterms (\ref{Deltambasis}) to  $s_\Ga + r^ {(1)} \delta {\cal
  S}$-symmetric operators, 
 we will 
proceed in two steps. First, the counterterms are written as
$s_\Gamma$-symmetric operators, and second they are extended to $s_\Ga
+ r^ {(1)} \delta {\cal S}$-symmetric ones.

The three field redefinitions $\Ga^ {(l)}_{{\rm ct}, \varphi}(\xi) ,
\Ga^ {(l)}_{{\rm ct}, \psi}(\xi)$ and
$\Ga^ {(l)}_{{\rm ct}, \psi\varphi} (\xi)$ in (\ref{Deltambasis}) 
have already been
constructed as $s_\Ga$-symmetric operators in
(\ref{ctvarphi}), (\ref{ctpsi}) and (\ref{ctpsivarphi}):
\begin{eqnarray}
\label{ctphisym}
\Ga^ {(l)}_{{\rm ct}, \phi}(\xi) &  = & {\cal N}_\phi ^ {(l)} \Gacl+
\Delta_{Y,\phi}^ {(l)} \qquad \mbox{with} \quad \phi =
\varphi,\psi,\psi\varphi\ .
\end{eqnarray}
 A straightforward
calculation shows that the operators ${\cal N}_\varphi ^ {(l)} $ and ${\cal N}_\psi ^ {(l)} $ are 
$s_\Ga + r^ {(1)}\delta
{\cal S}$-invariant operators at all
loop orders $l$.
The local invariant $\Ga^ {(l)}_{ct, \psi\varphi}$ can be extended to an
 $s_\Ga + r^ {(1)}\delta
{\cal S}$-invariant in a trivial way by constructing it
as an  $s_\Ga + r^ {(1)}\delta
{\cal S}$-variation of the corresponding local field monomial:
\begin{eqnarray}
 & & (s_\Ga + r^ {(1)} \delta {\cal S}) 
\intd \sqrt 2  f^ {(l)}_{\psi\varphi}(\tilde \xi)
\bigl(\chi^ {E^ {(2l)}} \varphi_L Y_{\psi_L}  + \chibar^  {E^ {(2l)}}
\phibar_L Y_{\psibar_L} + (_{L\to R})\bigr)  \\
& = & {\cal N}^ {(l)}_{\psi\varphi} \Ga + \Delta^ {(l)}_{Y, \psi\varphi} +
 r ^{(1)}
\delta \Delta^ {(l+1)}_{Y, \psi\varphi} \nonumber
\end{eqnarray}
where 
\begin{equation}
\label{ctpsiphisym}
\delta \Delta^ {(l+1)}_{Y, \psi\varphi} = -  8 l \intd \sqrt 2 e^ {2(l+1)}
f^ {(l)}_{\psi\varphi}(\tilde \xi) (Y_{\psi_L} \sigma^ \mu \epsilonbar + 
\epsilon \sigma^ \mu Y_{\psibar_l} + (_{L\to R}))V_\mu   \ .
\end{equation} 
Therefore ${\cal N}_\varphi , {\cal N}_\psi$ and ${\cal N}_{\psi\varphi}$
are symmetric operators describing the gauge-depen\-dent field
renorma\-lizations in the CS equation. 

The  operator which expresses  the field monomial $ \Ga_{\rm ct, kin}$ in (\ref{Deltambasis})
 is  a simple extension of the expression with constant
coupling (\ref{Dkine}):
\begin{eqnarray}
\Ga^ {(1)}_{\rm ct, kin} &= &-\frac 12 \intd e^ 2\Bigl(
e \frac{\delta} {\delta e} - A^ \mu   \frac{\delta} {\delta A^ \mu}
- \lambda^ \alpha  \frac{\delta} {\delta \lambda^ \alpha} - 
\lambdabar^ \alphadot  \frac{\delta} {\delta \lambdabar^ \alphadot}
\nonumber 
\\ & & \phantom{  -\frac 12 \intd e} 
+ Y_ \lambda^ \alpha  \frac{\delta} {\delta Y_\lambda^ \alpha} + 
Y _{\lambdabar}^ \alphadot  \frac{\delta} {\delta Y _{\lambdabar}^ \alphadot}
- c  \frac{\delta} {\delta c} \nonumber 
\\ & & \phantom{  -\frac 12 \intd e} 
+ B  \frac{\delta} {\delta B} +  \cbar\frac{\delta} {\delta \cbar} 
 - 2(\xi (x) + \xi)  \frac{\delta}
{\delta \xi} - 2 \chi_\xi  \frac{\delta} {\delta \chi_\xi}\Bigr)\;\Gacl
\nonumber \\
& & -\frac 12 \intd s (e^ 2) \bigl(\lambda^ \alpha Y_{\lambda \alpha}+
\lambda_ \alphadot Y_{\lambdabar}^ { \alphadot}\bigr) \nonumber \\
& \equiv & - \frac 12 {\cal D}_{\rm kin} \Gacl + 
\intd s (e^ 2) \bigl(\lambda^ \alpha Y_{\lambda \alpha}+
\lambda_ \alphadot Y_{\lambdabar}^ { \alphadot}\bigr)\ .
\label{Dkin}
\end{eqnarray}
As can be immediately verified, ${\cal D}_{\rm kin}  $
is
a $s_\Ga + r ^{(1)}\delta {\cal S}$-symmetric operator.

Unlike $\Ga_{\rm ct, kin}$, the operator expression of
 $\Ga_{\rm ct, \rm matter}$ cannot be obtained as a simple extension
 of the operator with  constant coupling (\ref{ctmattere}). As already
 mentioned in section 6, at this point the axial vector multiplet enters
 the construction.
Indeed, $\Ga_{\rm ct, \rm matter}$ can 
be decomposed  into field redefinitions of the matter fields
(\ref{ctvarphi}), (\ref{ctpsi}), (\ref{ctpsivarphi}) and  a
new $s_{\Gamma_{\rm cl}}$-invariant counterterm $\Ga
^{(l)}_{{\rm ct},Vv}$:
\begin{eqnarray}
\label{MatterCt}
\Ga^ {(l)}_{\rm ct, matter} &= &\frac 12  
\Bigl(\Ga^{(l)}_{\rm ct,\varphi}(f_\varphi(\xi)= 1)+\Ga^{(l)}_{\rm ct,\psi}
(f_\psi(\xi)= 1)+\Ga^{(l)}_{\rm ct,\psi\varphi}
(f_{\varphi\psi}(\xi)= 1) \nonumber
\\ & & { } \phantom{\frac 12} + \Ga^ {(l)}_{{\rm ct},Vv}\Bigr)\ .
\end{eqnarray}
This new invariant counterterm  describes field redefinitions of the axial
vector multiplet into the components of the local supercoupling $E^
{2l}$ 
and field redefinitions of 
the $\mathbf q$ and $\mathbf {\qbar}$-multiplets:
\begin{eqnarray}
\Ga ^{(l)}_{{\rm ct},Vv}
& = & \intd
 \biggl( v^ {{(E^{2l})} \mu} \frac {\delta} {\delta V^ \mu} +
\lambda^{{(E^{2l})}\alpha}
\frac {\delta} {\delta \lambdaV^\alpha}
+
\lambdaVbar ^ {{(E^ {2l})}\alphadot}
\frac {\delta} {\delta \lambdaVbar^\alphadot}
 \nonumber \\
& & { } \phantom{\intd}
+ d^ {(E^{2l})}  
\frac {\delta}{\delta \DV}
- i (\epsilon ^ \alpha \chi^ {(E^{2l})}_\alpha - \chibar^ {(E^{2l}) }
_\alphadot \epsilonbar^ \alphadot ) \frac {\delta}{\delta \tilde c} 
\nonumber \\
& & { }\phantom{\intd}
- 2 e^ {2l} (q +m) \frac {\delta }{\delta q} - 2 \bigl(e^{2l}
q^\alpha + 2 \chi^{{(E^{2l})}\alpha}(q+m)\bigr) 
\frac {\delta} {\delta q^ \alpha}
 \nonumber \\
& & { } \phantom{\intd}  -2
\bigl(e^ {2l}q_F + 2 f^ {(E^{2l})}(q+m)-  \chi^{{(E^{2l})}\alpha} q_\alpha\bigr)
 \frac {\delta } {\delta q_F}
 \nonumber \\
& & { } \phantom{\intd}
- 2 e^ {2l} (\qbar +m) \frac {\delta }{\delta \qbar} - 2 \bigl(e^{2l}
\qbar^\alphadot +2 \chibar^{{(E^{2l})}\alphadot}(\qbar+m)\bigr) 
\frac {\delta} {\delta \qbar^ \alphadot}
 \nonumber \\
& & { } \phantom{\intd} -2
\bigl(e^{2l}\qbar_F
 + 2\fbar^ {(E^{2l})}(\qbar+m)-  \chibar^{(E^{2l})}_\alphadot \qbar
 ^\alphadot\bigr) 
 \frac {\delta } {\delta \qbar_F} \biggr)\Ga_{\rm cl} \nonumber \\
& \equiv&  {\cal D}^{(l)}_{Vv} \Gacl \ .
\label{DVv}
\end{eqnarray}

In eq.~(\ref{DVv})  the components of the  multiplet $E^
{2l}$ are defined by the following expansion:
\begin{eqnarray}
\label{E2ldef}
E^{2 l}(x, \theta, \thetabar )  & = & \bigl({\mbox{\boldmath{$\eta$}}}(x,
\theta, \thetabar ) +{\mbox{\boldmath{$\etabar$}}}(x, \theta,
\thetabar )\bigr)^ 
{-l} \nonumber  \\
&\equiv &  e^{2l} (x) + \theta^ \alpha \chi^{(E^ {2l})}_\alpha +
 \chibar^{(E^{2 l})}_\alphadot \thetabar^ \alphadot + 
\theta^ 2 f^ {(E^ {2l})} +\thetabar^ 2 \fbar^ {(E^ {2l})} 
+ \theta \sigma^ \mu \thetabar v_\mu^ {(E^ {2l})} 
\nonumber  \\
& & { }
+i \theta^ 2 (\lambdabar^{(E^ {2l})}+
\frac 12 \partial_\mu \chi  \sigma^\mu)
 \thetabar 
-i \thetabar^ 2 \theta (\lambda^{(E^ {2l})} +
\frac 12  \sigma^\mu \partial_\mu \chibar^{(E^{2l})} )
\nonumber  \\
& & { }
   + \frac 14
\theta^ 2 \thetabar^2  (d^ {(E^ {2l}) } - \Box e^{2l}) 
\end{eqnarray} 
with the explicit form of the lowest components:
\begin{eqnarray}
\chi^{(E^ {2l})}_\alpha & = & - l e^ {2(l+1)} \chi_\alpha \ ,
\qquad
\chibar^{(E^ {2l})}_\alphadot = - l e^ {2(l+1)} \chibar_\alphadot\ ,
 \nonumber \\
v_\mu^ {(E^ {2l})} &  = & i\;l e^ {2(l+1)} \partial_\mu (\eta - \etabar)  
+ \frac 12 l (l+1) e ^ {2(l+2)}( \chi \sigma_\mu \chibar).
\label{vE2l}
\end{eqnarray}
The explicit form of the higher components is not relevant for the further
 construction. It is sufficient to know their supersymmetry transformations,
 which are uniquely determined by the expansion in superspace (see
 (\ref{susyreal}) in appendix A).

The operator ${\cal D}_{Vv}$  is  $s_\Ga$-invariant, but it
does not commute with
the anomalous Slavnov--Taylor operator.  A lengthy but straightforward
calculation 
shows that it can be extended 
to an $s_\Ga + r^ {(1)}\delta 
{\cal S}$-invariant operator by including redefinitions of the local
 coupling and an anomalous dimension for the axial vector multiplet
 into the operator.
Their coefficients are uniquely determined,
and one finds the following   $s_\Ga + r^ {(1)}\delta 
{\cal S}$-invariant operator:
\begin{equation}
\label{DvVsym}
{\cal D}^ {\rm sym}_{Vv} \equiv
{\cal D}^ {(l)}_{Vv} - r^ {(1)}\bigl( 4 {\cal D}^ {(l+1)}_{e} +
8 l ( {\cal N}^ {(l+1)}_V - 8 (l+1)r ^{(1)} \delta {\cal N}^ {(l+2)}_V )
 \bigr) .
\end{equation}
In this expression
  ${\cal N}^ {(l+1)}_V $ contributes to an anomalous dimension of the
axial vector field and its superpartners:
\begin{eqnarray}
{\cal N}^ {(l)}_V & = &
 \intd \biggl( e^ {2l}\Bigl( V^ {\mu} \frac {\delta }{\delta V^ \mu} +
 \lambdaV^ {\alpha} \frac {\delta }{\delta \lambdaV^ \alpha} +
 \lambdaVbar^ {\alphadot} \frac {\delta }{\delta \lambdaVbar^ \alphadot} +
 \tilde D \frac {\delta }{\delta \tilde D}\Bigr) \nonumber \\
& & \qquad -\frac i 2 V_\mu (\sigma^ \mu \chibar^ {(E ^{2l})})^ \alpha \frac {\delta }{\delta
 \lambdaV^ \alpha} + \frac i2  V_\mu (\chi^  {(E ^{2l})} \sigma^ \mu )^ \alphadot       
\frac {\delta }{\delta \lambdaVbar^ \alphadot}  \nonumber \\
& &  \qquad + 2 \Bigr( V^ \mu v_{\mu}^
 {(E^ {2l})}  + i \lambdaV^ \alpha \chi^ {(E^ {2l})}_\alpha 
- i\chibar^ {(E^ {2l})}_ \alphadot\lambdaVbar^ \alphadot 
 \Bigl)
\frac {\delta }{\delta \tilde D }  \biggr),
\end{eqnarray}
and 
\begin{eqnarray}
\delta {\cal N}^ {(l)}_V
 & = &  \intd e^ {2l} V^ \mu V_\mu \frac {\delta }{\delta \tilde D }\ ;
\end{eqnarray}
 the operator $ {\cal D}_e $ describes a
 redefinition of the coupling $e(x)$ and its superpartners:
\begin{eqnarray}
\label{De}
{\cal D}^ {(l+1)}_e & = & 
 \intd \biggl( e^ {2l+3}\frac {\delta }{\delta e} - 2 \bigl(
\chi^ {(E^ {2l})\alpha} \frac {\delta }{\delta \chi^ \alpha} +
\chibar^ {(E^ {2l})\alphadot} \frac {\delta }{\delta \chibar^
\alphadot}\bigr) \nonumber \\
& & \phantom{ 4 \intd} - 2
\bigl(
f^ {(E^ {2l})} \frac {\delta }{\delta f} +
\fbar^ {(E^ {2l})} \frac {\delta }{\delta \fbar}\bigr) \biggr)
\nonumber \\
& & \phantom{ 4\intd}
 - e^{2(l+1)}\Bigl( A^ \mu   \frac{\delta} {\delta A^ \mu}
+ \lambda^ \alpha  \frac{\delta} {\delta \lambda^ \alpha} +
\lambdabar^ \alphadot  \frac{\delta} {\delta \lambdabar^ \alphadot}
\nonumber \\ 
& & \phantom{4\intd - e^ {2(l+1)}}
- Y_\lambda^ \alpha  \frac{\delta} {\delta Y_\lambda^ \alpha} -
Y _\lambdabar^ \alphadot  \frac{\delta} {\delta Y_\lambdabar^ \alphadot}
+ c  \frac{\delta} {\delta c} \nonumber \\
& & \phantom{4\intd - e^ {2(l+1)}}
- B  \frac{\delta} {\delta B} -  \cbar\frac{\delta} {\delta \cbar} 
 + 2(\xi (x) + \xi)  \frac{\delta}
{\delta \xi} + 2 \chi_\xi  \frac{\delta} {\delta \chi_\xi}\Bigr)\ .
\end{eqnarray}

Thus, we have  expressed all five invariant basis elements of
  $\Delta^ {(l)}_{m}$ (\ref{Deltambasis}) in terms of the
  $s_\Ga + r^
{(1^)}\delta {\cal S}$ symmetric  operators ${\cal D}_{\rm kin}$
  (\ref{Dkin}),  $
{\cal D}^ {\rm sym}_{Vv}$ (\ref{DvVsym}) and  $
{\cal N}_\varphi , {\cal N}_\psi,{\cal N}_{\psi\varphi}$
(\ref{ctphisym}).
Therefore the dilatational anomalies can be absorbed into the following
Callan-Symanzik operator
\begin{eqnarray}
{\cal C}& = & \mu_i D _{\mu_i} +
\hat \beta_e^ {(1)} {\cal D}_{\rm kin}+ 
\sum_{l} \bigl(4 r ^{(1)} \hat \gamma^ {(l)} {\cal D}^ {(l+1)}_e
 - \hat \gamma ^ {(l)} {\cal D}^ {(l)}_{Vv} + 8
r ^{(1)}\hat \gamma^ {(l)} (l {\cal N}_V ^{(l+1)}\nonumber \\
& & { }\quad  - 8 r ^{(1)}
l (l+1)\delta {\cal N}_V^{(l+2)})
- \hat\gamma_\varphi^ {(l)} {\cal N}_\varphi^ {(l)} 
- \hat \gamma_\psi^ {(l)} {\cal N}_\psi^ {(l)} -
\hat \gamma_{\psi\varphi}^ {(l)} {\cal N}_{\psi\varphi}^ {(l)}
 \bigr)
\end{eqnarray}
and the algebraic construction
yields the CS equation of SQED with local coupling and gauged axial symmetry
\begin{equation}
\label{CS}
{\cal C}\Ga = \Delta_Y \ .
\end{equation}

The two main properties are the restriction of  ${\cal D}_{\rm kin}$
to one-loop and the connection between ${\cal D}_{e}$, ${\cal D}_{Vv}$ 
and ${\cal N}_V$ via the anomaly coefficient $r^ {(1)}$.
Thus there are only two gauge-independent, physical coefficients
in the  
CS equation of SQED with local coupling and gauged axial symmetry, 
namely the one-loop $\beta$-function $\hat\beta_e^ {(1)}$ and the anomalous
dimension  $ \hat \gamma^ {(l)}$.  

\subsection{The limit to constant coupling}

In order to  clarify the significance of the CS coefficients in (\ref{CS})
 we
 turn to the limit of constant coupling and constant gauge
 parameter. In this limit all higher components in the
 $\theta$-expansion of the supercoupling $E^ {2l}$ vanish. Therefore
 the only parts which are left from the operator ${\cal D}^
 {\rm sym}_{Vv}$  (\ref{DvVsym}) are contributions to the anomalous
 dimension of $q$-fields and of the axial vector multiplet, and a
 contribution to the gauge $\beta$-function. The latter
 arises from the operator ${\cal D}^ {(l+1)}_e$ (\ref{De}), and
together with 
the one-loop 
operator $ {\cal D}_{\rm kin}$
(\ref{Dkin})  it  determines the $\beta$-function $\beta_e$ of SQED:
\begin{equation}
\lim _{E\to e} (\hat \beta_e^ {(1)} {\cal D}_{\rm kin}+4 r^ {(1)} \sum_{l}
\hat \gamma^ {(l)} {\cal D}^ {(l+1)}_e ) \Ga = 
e ^3 (\hat \beta_e^ {(1)} + 4 r^ {(1)}  \gamma) \partial _e \Ga  + \cdots  
\end{equation}
and therefore
\begin{equation}
\label{betae}
\beta_e =   e ^2 (\hat \beta_e^{(1)} + 4 r^ {(1)} \gamma) 
\quad \mbox{\rm with} \quad \gamma = \sum_l e^ {2l} \hat \gamma^{(l)} \ .
\end{equation}
Inserting the values for the one-loop $\beta$-function and for the anomaly
coefficient (\ref{r1}),
\begin{equation}
\hat \beta^ {(1)}_e = \frac 1 {8\pi^2 } \ , \qquad r^ {(1)} = - \frac 1
{16\pi^2 },
\end{equation}
we find the closed expression for the $\beta$-function 
\begin{equation}
\beta_e = \hat \beta_e^{(1)} e ^2 (1 - 2 \gamma).
\end{equation}

In the limit to constant coupling the functional $\Ga$ 
in (\ref{CS}) is the generating functional of 1PI Green functions
 of ordinary SQED
extended by the  anomalous axial symmetry.
If  the axial vector field and its superpartners are set to zero in
addition 
 it 
coincides with $\Ga^ {\rm SQED}$.
Therefore we find in the limit to constant coupling from (\ref{CS})
the CS equation of
SQED with gauged axial symmetry, whose coefficients are 
restricted by the multiplet structure of the SQED action:
\begin{eqnarray}
\label{CSconst}
& & \Bigl(\mu_i\partial_{\mu _i} +
e^2(\hat \beta^ {(1)}_e + 4 r^ {(1)} \gamma) (e\partial_e - N_A - N_\lambda + N_{Y_\lambda} - N_c + N_B +
N_\cbar  - 2 \xi \partial_\xi)\nonumber \\
& & + 4 r^ {(1)}e^ 3 \partial_e \gamma (N_V + N_\lambdaV + N_{\tilde D})
+2  \gamma (N_q + N_{q^ \alpha} + N_{q_F} )  \nonumber \\
& & -  \gamma _\varphi (N_{\varphi_L} + N_{\varphi_R} - 
N_{Y_{\varphi_L}} - N_{Y_{\varphi_R}}) 
-  \gamma _\psi (N_{\psi_L} + N_{\psi_R} - 
N_{Y_{\psi_L}} - N_{Y_{\psi_R}})  \Bigr)\, \Ga \nonumber \\
& &  =
(1-2\gamma) \intd m \bigl(\frac{\delta}{\delta q} +
\frac{\delta}{\delta \qbar}
\bigr)\, \Ga + \Delta_{Y,\psi \varphi}\ .
\end{eqnarray}

In the CS equation the non-renormalization theorems are identified as
the various restrictions on the CS coefficients: The non-renormalization
of chiral vertices    (cf.~(\ref{Gactinv})
and section 5.1) is expressed by
  the common anomalous dimension $\gamma$ for the mass 
and the $q$-field. However, non-renormalization in the strict sense is covered up by
the gauge dependent anomalous dimensions  $\gamma_\psi$ and
$\gamma_\varphi$ of the matter fields:
\begin{eqnarray}
\gamma_\varphi &  = & \sum _l e^ {2l} f^ {(l)}_\varphi (\xi)\hat \gamma_\varphi^
 {(l)}\ ,
\qquad \quad
\gamma_{\psi}  =   \sum _l e^ {2l} f^ {(l)}_\psi (\xi)\hat 
\gamma_\psi^ {(l)}\ .
\end{eqnarray}
Only in the supersymmetric gauge they turn out to be gauge independent and to
coincide with the anomalous dimension of the mass, making there the
non-renorma\-lization theorems of chiral vertices manifest.

The closed form of the gauge $\beta$-function (\ref{betae}) is
a consequence of
the non-renormalization of the photon self energy in loop orders $l \geq 2$
 (cf.~(\ref{Gactinv})
and section 5.2).  An additional interesting result of the present
construction
is the anomalous dimension of the axial vector multiplet appearing in
(\ref{CSconst}),
\begin{equation}
\label{gaV}
 \gamma_V = - 8 r ^{(1)}  \sum _l l \hat \gamma^ {(l)}  e^ {2(l+1)} 
 = -4 r^{(1)} e^ 3\partial_e \gamma.
 \end{equation}
 It  illustrates  the deep interplay 
between local couplings and
gauged axial symmetry.

With the use of local couplings and an axial vector field we have
derived all improved properties of the CS
equation in the Wess--Zumino gauge. Previously, the corresponding expressions 
have only been accessible in the manifestly supersymmetric gauge and
by the construction of the supercurrent. Even there the explicit terms for
the anomalous dimension of the axial current are available only with
quite some technical effort.  For this reason, we are convinced that  
the technique  of local couplings can also enlighten some
further unproven renormalization properties in more complex theories.

\section{Supersymmetry breaking of the axial current Green functions
and the Konishi anomaly}

In section 5.2 we have shown that the photon self energy in $l\geq 2$ 
is completely
 determined by non-local Green functions (see  (\ref{Gaphoton}) and
(\ref{GachilambdaA})). Although no axial-current Green
functions are involved, the expressions we have 
 obtained resemble the axial anomaly, because they
 are divergent and ask for regularization, but  their divergent part 
is completely  determined by the non-local part via gauge invariance.
On  the other hand, the axial anomaly has played an important role in
the algebraic construction of the CS equation. Using this construction the gauge
 $\beta$-function (\ref{betae}) has been shown to be completely determined by
the anomalous dimension $\gamma$ and the coefficient of the axial anomaly.
 In this section, an underlying connection between these results
  is revealed by deriving
explicit relations between the
photon self energy and axial-current Green functions.

As the first relation, the high-energy logarithms of the 2-loop photon
self energy are completely expressed by one-loop axial-current Green
functions. 
%
By (\ref{Gaphoton}), the 2-loop photon self energy is completely
determined by vertex functions of the form $\Ga_{\chi\lambda A^\mu}$,
and these vertex functions are in turn determined by (\ref{GachilambdaA}),
\begin{eqnarray}
\label{GachilambdaA2}
2 e ^2 l \Ga^ {(l)}_{\chi_\gamma \lambda^ \beta A^ \mu  } (-p , p)
& = & 
i \sigma ^ \rho _{\beta \betadot} \Ga^ {(l)}_{\chibar_\betadot \chi_ \gamma
A^ \mu A ^\rho} (p,-p) - p_\rho (\sigmabar^ {\rho \mu})^ \alphadot{ }_\betadot
\Ga^ {(l)}_{\chibar_\betadot \chi_ \gamma
\lambda ^ \beta  \lambdabar ^ \alphadot} (- p, p) \nonumber \\
& & - \sum_{k= 0}^ l
\Ga^ {(l-k)}_{\epsilonbar^ \betadot
\chibar_\betadot \chi_ \gamma
A^ \mu Y_{\lambdabar}^ {\alphadot}} (p, -p) \Ga^ {(k)}_{\lambda^ \beta
\lambdabar_\alphadot } ( -p , p),
\end{eqnarray}
in terms of vertex functions with a $\chibar \chi$-insertion.
All such Green functions vanish at the 0- and 1-loop level.
In order to find the high-energy behaviour, we apply the CS equation
(\ref{CS}) on the right-hand-side of (\ref{GachilambdaA2}).
At 2-loop order the  only non-vanishing term  comes from the
 operator ${\cal D}^ {(1)}_{vV} = \int v^ {(E^2)\mu}\delta _{V^ \mu} +
 \cdots$ (\ref{DVv}). This operator is the crucial element for
 relating the Green functions $\Ga_{\chibar^ 
  \betadot \chi^ \gamma \ldots}  $ 
to Green functions with an axial current insertion at zero momentum,
for it describes a field renormalization of
the axial
vector field into the vector component $v_\mu^
{(E^ {2l})} 
 $  of the local supercoupling (\ref{vE2l}), which  contains the
 combination $\chi \sigma \chibar$.
 Explicitly we find
in  two-loop order:
\begin{equation}
\label{CSphotondirekt}
m\partial_m 4 e^2
\Ga^ {(2)}_{\chi^ \gamma \lambda^ \beta A^\mu} = e ^ 6
\hat\gamma^ {(1)} \sigma^\rho_{\gamma \betadot}  {\widetilde \Ga}^ {(1)}_{\rho \mu}{ }^ 
 \betadot{}_{\beta}(p,-p) + {\rm soft\ terms}
\end{equation}
with 
\begin{eqnarray}\label{Gatilde}
  {\widetilde \Ga}_{\rho \mu
 \beta\betadot}(p,- p) & = &
- i \sigma ^ {\lambda}_{\beta \betadot}   \Ga_{V^ \rho
A^ \mu A ^\lambda}
 (0,p,- p) + p_{\lambda} (\sigmabar^ {\lambda \mu})^ \alphadot{ }_ { \betadot} 
\Ga_{V^\rho 
\lambda ^ \beta  \lambdabar ^ \alphadot} (0,- p, p) \nonumber \\
& & +
\Ga_{\bar \epsilon^ \betadot
V^ \rho
A^ \mu Y_{\lambdabar}^ {\alphadot}} (0,p,-p) \Ga_{\lambda^ \beta
\lambdabar_\alphadot } (- p , p)
 \ .
\end{eqnarray}
Via eq.\ (\ref{Gaphoton}), the identity (\ref{CSphotondirekt})
expresses the high-energy behaviour of the photon self energy
explicitely in terms
of lower order Green functions with axial-current insertions
${\widetilde  \Ga}_{\rho \mu \betadot\beta}(p,-p)$.  
Therefore, all 2-loop high-energy logarithms of the photon self energy
are due to subdivergences related to one-loop triangle diagrams.

It is possible to derive a complementary relation for the photon self
energy, where the axial Green functions in the combination of
eq.~(\ref{Gatilde}) are  entirely expressed
 by the photon self energy.
From
\begin{equation}
\frac{\partial}{\partial \epsilonbar^ \betadot} 
\frac{\delta}{\delta V^ \rho(z)} \frac{\delta}{\delta A^ \mu(x)} 
\frac{\delta}{\delta \lambda^ \beta(y)} ({\cal S} + r^ {(1)} \delta
{\cal S}) (\Ga) = 0
\end{equation}
one gets the relation 
\begin{eqnarray}
\label{KonishiWZ}
 {\widetilde \Ga}_{\rho \mu
 \betadot\beta}(p,- p) 
& = &
  8 r^ {(1)} \sigma^ \rho _{\alpha \betadot} 
\Ga_{\chi_\alpha A^\mu \lambda^ \beta}( p,- p)
\end{eqnarray}
between the axial Green functions in
$ {\widetilde \Ga}^ {(l)}_{\rho \mu
 \betadot\beta}(p,-p)   $ and the vertex function $\Ga_{\chi \lambda
 A^\mu}$.
By using the relation
(\ref{Gaphoton}) to eliminate $\Ga_{\chi  A^\mu \lambda}$, the
function   ${\widetilde  \Ga}_{\rho \mu 
 \betadot\beta}(p,-p)$ can be related to the photon self energy:
\begin{equation}
\label{KonishiWZ0}
p^ \lambda  (\sigmabar^ \rho \sigma_ {\mu \lambda} )^ {\betadot\beta}
{\widetilde \Ga}^ {(l)}_{\rho \nu\beta \betadot} (-p, p) -
p^\lambda  (\sigmabar^ \rho \sigma_ {\nu \lambda} )^ {\betadot \beta}
{\widetilde \Ga}^ {(l)}_{\rho \mu\beta \betadot} (p,- p) = 64 (l-2)r^
{(1)}\Ga^ {(l-1)}_{A^ \mu A^ \nu} .
\end{equation}
This is the announced relation. It can be solved for the photon self
energy in terms of axial-current Green functions at all orders,
except for one-loop order. This exception corresponds to the only independent
divergent contribution to the photon self energy in one loop order. 

The two relations we have derived can be combined to rearrange the CS equation
(\ref{CSphotondirekt})  into an ordinary scaling equation for the
vertex function $\Ga_{\chi \lambda A}$ or, using (\ref{Gaphoton}), for the
photon self energy. Inserting (\ref{KonishiWZ0}) into
(\ref{CSphotondirekt})
and using the relation (\ref{Gaphoton}) we find the CS equation
of the photon self energy in 2-loop order, which is in agreement with
the algebraic construction of section 8 (cf.~in particular 
eq.~(\ref{CSconst})):
\begin{equation}
\label{CSphoton}
m\partial_m \Ga^ {(2)}_{A^ \mu A^\nu} = 8  e ^2 r^ {(1)} \gamma^ {(1)} 
\Ga^ {(0)}_{A^ \mu
A^ \mu} + {\rm soft\  terms} \ .
\end{equation}


Eq.\ (\ref{KonishiWZ}) relates the vertex
function $\Ga_{\chi\lambda A^\mu}$ to Green functions with 
axial-current  insertions, and eq.\ (\ref{GachilambdaA2}) relates the
same vertex function to the Green functions with $\chi \chibar$-
insertions,
multiplied with the anomaly coefficient $r ^{(1)}$. Comparing both
identities reveals a one-to-one 
correspondence of the Green functions $\Ga^ {(l)} _{V^\mu\ldots}$
and $r^{(1)}\Ga^ {(l-1)}_{\chibar^ \betadot \chi^ \gamma \ldots}$.
This is an underlying connection between the axial-current Green
functions and the non-local expressions defining the 
photon self energy. 

An interpretation of the identity (\ref{KonishiWZ}) is the invariance
of axial Green functions under supersymmetry transformations. In the
anomalous Slavnov--Taylor identity, the supersymmetry
transformations of the axial multiplet contain contributions from the
axial anomaly, and correspondingly (\ref{KonishiWZ}) relates axial
Green functions to Green functions of lower order, multiplied with the
anomaly coefficient $r^{(1)}$. From the point of view of the
unmodified supersymmetry transformations without the anomaly
contributions, (\ref{KonishiWZ}) can be viewed as an expression for
a supersymmetry breaking for axial-current Green functions, induced
by the Adler-Bardeen anomaly in the Wess--Zumino gauge.

Furthermore, the  expressions (\ref{KonishiWZ}) and
(\ref{KonishiWZ0}) have an  analogue in
superspace. In the  supersymmetric gauge  
 the axial Ward identity is covariantly decomposed  into chiral
and antichiral transformations which read \cite{CPS79}:\footnote{We use the
conventions  of Clark, Piguet and Sibold \cite{CPS79} with 
minor 
modifications, which concern the massive photon field and  the
non-renormalization of chiral vertices. We only give a short sketch of
the construction and refer for details to the original paper.}
\begin{eqnarray}
\label{waxialsusy}
\bar{D} \bar{D} I_5\;  \Ga & = & ( A_+ \frac {\delta} {\delta A_+} + A_- \frac
{\delta} {\delta 
A_-})\Ga + \frac 12 m [A_+ A_-]_2 \cdot \Ga  + r^ {(1)} [W^ \alpha
W_\alpha]_3\cdot \Ga , \nonumber \\
 {DD} I_5 \;\Ga & = & (\bar A_+ \frac {\delta} {\delta \bar A_+} + \bar
A_- \frac {\delta} {\delta 
\bar A_-})\Ga + \frac 12 m [\bar A_+ \bar A_-]_2 \cdot \Ga + r^ {(1)}
[
\bar W_ \alphadot \bar W^
\alphadot]_3\cdot \Gamma ,
\end{eqnarray}
where $W_\alpha $ and $\bar W_\alphadot$ are the supersymmetric field
strength tensors of the real superfield containing the photon and
photino
\begin{equation}
W_\alpha = - \frac 1 8 {\bar D \bar D} D_\alpha \phi,
\qquad
\bar W_\alphadot = - \frac 1 8   DD {\bar D}_\alphadot \phi,
\end{equation}
and $I_5$ is the gauge invariant multiplet of the matter Lagrangian:
\begin{equation}
I_5 = \frac 1 {16} [\bar A_+ e^ {e\phi} A_+ + 
\bar A_- e^ {- e\phi} A_- + {\cal O} (\hbar)]_2\ .
\end{equation}
Like (\ref{KonishiWZ}) and (\ref{KonishiWZ0}), these identities
represent a supersymmetric extension of the anomaly found  here
in the 
decomposition of the anomaly into a chiral and antichiral part. 

 Subtraction of the chiral and antichiral
identities in (\ref{waxialsusy}) yields the axial Ward identity with
its anomaly. However, if one integrates and sums the two identities,
one obtains an identity for the matter insertion $\intV I_5$. It
contains a field redefinition and a mass insertion, but involves
in particular also insertions of
 the gauge invariant kinetic term  $\intS W^ \alpha W_\alpha$ and its
complex conjugate. This contribution is due to the
Adler--Bardeen 
anomaly of the axial current.
 (The identity of the matter term insertion
 is sometimes called Konishi
anomaly \cite{KO81}.) In combination with the local version of 
the CS equation it yields the closed
form of the $\beta$-function in the usual superspace approach without
local couplings.

In the Wess--Zumino gauge 
 the anomalous axial Ward identity (\ref{WIaxialanomalous})
  is left as the only equation 
of  the covariant expressions in superspace, but the Adler--Bardeen
 anomaly induces a supersymmetry breaking of axial current Green functions. 
And the relations (\ref{KonishiWZ}) and (\ref{KonishiWZ0}) 
we have derived from  the anomalous Slavnov--Taylor identity have a
 similar physical content
 as the Konishi anomaly in superspace, since they relate axial
current Green functions to the photon self energy.
In the Wess--Zumino gauge they can therefore be considered as the
analogue to the Konishi anomaly in the manifestly
supersymmetric gauge.

\section{Conclusions} 

We have traced back the non-renormalization theorems of chiral
vertices and the generalized non-renormalization theorem of the photon
self energy in SQED on a common algebraic origin, namely the nature of
supersymmetric Lagrangians as the highest components of
supermultiplets.
This fact has been exploited using the extension of the gauge
coupling to an external  superfield.
 On the basis of symmetric
counterterms, the non-renormalization theorems appear as absence of
independent counterterms to the chiral vertices in $l \geq 1$ and  to the kinetic term of the photon multiplet
in loop order $l\geq  2$.
We have related the corresponding vertex functions to non-local
expressions. There is a remarkable difference between the usual and the
generalized 
non-renormalization theorem: Chiral vertex functions are --- up
to gauge dependent field redefinitions ---
related to superficially convergent Green functions, whereas  the
photon self energy is related to linearly divergent Green functions.
The latter expressions turn out to be independent of the
renormalization procedure only if gauge invariance is taken into
account. In this sense, the expressions determining the photon self
energy 
remind of  the triangle diagrams of  the axial current. 

Indeed, a deep connection between the local supercoupling and the
axial current 
is revealed when we complete SQED with local gauge coupling to a
multiplicatively renormalizable theory: It turns out to be necessary
to introduce an axial vector multiplet whose vector component couples
to the axial current. The extended action of SQED with the
external superfield of the local coupling and the axial vector
multiplet gives a complete description of all known renormalization
properties of SQED including the Adler--Bardeen anomaly. Owing to the
local gauge 
coupling the Adler--Bardeen anomaly can be absorbed into the
Slavnov--Taylor operator and the extended model can be renormalized
algebraically  by using the anomalous Slavnov--Taylor identity. Even
the non-renormalization theorem of the Adler--Bardeen anomaly is shown
to be an almost trivial consequence of the construction with local
coupling: Only the one-loop order of the anomaly is a
total derivative; higher orders would contribute to the global
integrated axial Ward identity, hence their coefficients  vanish.

All these properties together make possible to derive the
Callan--Symanzik equation for SQED including the anomalous axial
current from an algebraic construction with the anomalous
Slavnov--Taylor identity. The Callan--Symanzik equation 
depends only on two physical coefficients, the one-loop  gauge
$\beta$-function and the anomalous dimension of mass, $\gamma$. The
higher orders of the $\beta$-function and the anomalous dimension of
the anomalous axial current are related to these coefficients via the
anomaly coefficient. In the course of the construction also the
gauge-dependent anomalous dimensions of the matter fields are
identified and disentangled from the non-renormalization properties.

In summary, usual SQED appears in the extended construction
as a limiting theory of a more
fundamental one, which  includes all known
non-renormalization theorems in its structure. It is remarkable that
 the soft breakings of the Giradello--Grisaru class can be
introduced without further modifications. They are the lowest
components of the multiplets of gauge invariant Lagrangians and are as
such already included in the present construction. They can be
explicitely obtained by a shift in the highest component
of the corresponding external fields.

\vspace{1cm}
{\bf Acknowledgments}

We thank  R. Flume and W. Hollik for many
 valuable discussions and K. Sibold for a critical
 reading of the manuscript. 
E.K. is grateful to the Fachbereich Physik of the Universit\"at
Kaiserslautern for kind hospitality where  parts of this work have been
  done.

\begin{appendix}

\section{Conventions and Notations}

\paragraph{2-Spinor indices and scalar products:}

\begin{eqnarray}\label{DefEpsilon}
\epsilon_{\alpha\beta} & = & - \epsilon_{\beta\alpha},\quad
\epsilon_{12} =1,
\quad\epsilon^{\alpha\beta}\epsilon_{\beta\gamma}  =  \delta^\alpha _\gamma
,\\
\epsilon_{\dot\alpha\dot\beta} & = & 
- \epsilon_{\dot\beta\dot\alpha},\quad 
\epsilon_{\dot1\dot2} =
1,\quad\epsilon^{\dot\alpha\dot\beta}\epsilon_{\dot\beta\dot\gamma} 
=  \delta^{\dot\alpha} _{\dot\gamma} 
,\nonumber \\
\psi\chi & = & \psi^\alpha\chi_\alpha\ , 
\quad
\psi^\alpha  = 
\epsilon^{\alpha\beta}\psi_\beta\ ,
\nonumber \\
\overline\psi\overline\chi & = & 
\overline\psi_{\dot\alpha}\overline\chi^{\dot\alpha}\ ,
\quad
\overline\psi_{\dot\alpha}  = 
\epsilon_{\dot\alpha\dot\beta}\overline\psi^{\dot\beta}\ .
\end{eqnarray}

\paragraph{$\sigma$ matrices:}

\begin{eqnarray}\label{Paulimatrizen}
&&
\sigma^\mu_{\alpha\dot\alpha}  =  (\sigma^ 0_{\alpha\dot\alpha},
\sigma^1_{\alpha\dot\alpha},\sigma^ 2_{\alpha\dot\alpha},\sigma^
3_{\alpha\dot\alpha}) \ ,
\\
&&
\sigma^0  = \left(\begin{array}{cc}1&0\\0&1\end{array}\right),\quad
\sigma^1  = \left(\begin{array}{cc}0&1\\1&0\end{array}\right),\quad
\sigma^2  = \left(\begin{array}{cc}0&-i\\i&0\end{array}\right),\quad
\sigma^3  = \left(\begin{array}{cc}1&0\\0&-1\end{array}\right) ,
\nonumber\\
&&
\overline\sigma^{\mu\dot\alpha\beta}  =  \epsilon^{\beta \alpha} 
\epsilon^ {\dot \alpha \dot \beta} \sigma^ \mu_{\alpha \dot \beta} \ .
\nonumber \\&&
\label{DefSigmaMunu}
(\sigma^{\mu\nu})_\alpha\ ^\beta  = 
\frac{i}{2}(\sigma^\mu\overline\sigma^\nu-\sigma^\nu\overline\sigma^\mu)
_\alpha\ ^\beta
\ ,\quad
(\overline\sigma^{\mu\nu})^{\dot\alpha}\ _{\dot\beta}  = 
\frac{i}{2}(\overline\sigma^\mu\sigma^\nu-\overline\sigma^\nu\sigma^\mu)
^{\dot\alpha}\ _{\dot\beta}\ . \nonumber 
\end{eqnarray}

\paragraph{Complex conjugation:}

\begin{eqnarray}
(\psi\theta)^\dagger & = & \thetabar\psibar \ ,\\
(\psi\sigma^\mu\thetabar)^\dagger & = & \theta\sigma^\mu\psibar\ , \qquad
(\psi\sigma^{\mu\nu}\theta)^\dagger  =  \thetabar\sigmabar^{\mu\nu}\psibar\ .
\nonumber 
\end{eqnarray}

\paragraph{Derivatives:}

\begin{eqnarray}
\frac{\partial}{\partial\theta^\alpha} \theta^\beta 
& = & \delta_\alpha^\beta
\ ,\qquad
\frac{\partial}{\partial\theta_\alpha} \theta_\beta 
= -\delta_\beta^\alpha
\ ,\\
\frac{\partial}{\partial\thetabar^ \alphadot} \thetabar^ \betadot 
& = & \delta_\alphadot^ \betadot
\ ,\quad
\frac{\partial}{\partial\thetabar_\alphadot} \thetabar_\betadot 
= -\delta_\betadot ^{\alphadot}
\ . \nonumber 
\end{eqnarray}

\paragraph{Supersymmetry transformations in superspace}

\begin{itemize}
\item Superfields in the real representation
\begin{eqnarray}
\label{susyreal}
\delta_\alpha \phi(x, \theta, \thetabar) &  =  & \Bigl(
\frac{\partial}{\partial\theta^\alpha} + i \sigma^ \mu _{\alpha \dot
\alpha }\thetabar^ \alphadot \partial_\mu\Bigr) \phi (x, \theta,
\thetabar) \ ,  \\ 
\bar \delta_\alphadot \phi(x, \theta, \thetabar) &  = &  \Bigl( -
\frac{\partial}{ \partial\theta^\alphadot} - i \theta^ \alpha\sigma^ \mu _{\alpha \dot
\alpha } \partial_\mu\Bigr) \phi (x, \theta,
\thetabar)  \ . \nonumber 
\end{eqnarray}
\item Superfields in the chiral representation
\begin{equation}
\label{chiral}
\phi_{\rm  c}(x, \theta, \thetabar) = 
\phi(x + i \theta \sigma \thetabar, \theta, \thetabar) \ ;
\end{equation}
\begin{eqnarray}
\label{susychiral}
\delta_\alpha \phi_{\rm c}(x, \theta, \thetabar)  &  =   &
\frac{\partial}{\partial\theta^\alpha}  \phi_{\rm c} (x, \theta,
\thetabar) \ , \\
\bar \delta_\alphadot \phi_{\rm c}(x, \theta, \thetabar) &  =  & \Bigl( -
\frac{\partial}{ \partial\theta^\alphadot} - 2 i \theta^ \alpha\sigma^ \mu _{\alpha \dot
\alpha } \partial_\mu\Bigr) \phi_{\rm c} (x, \theta,
\thetabar) \ . \nonumber
\end{eqnarray}
\item Superfields in the antichiral representation
\begin{equation}
\label{antichiral}
\phi_{\rm ac}(x, \theta, \thetabar) = 
\phi(x - i \theta \sigma \thetabar, \theta, \thetabar) \ ;
\end{equation}
\begin{eqnarray}
\label{susyantichiral}
\delta_\alpha \phi_{\rm ac}(x, \theta, \thetabar) & = & \Bigl(
\frac{\partial}{\partial\theta^\alpha} + 
2 i \sigma^ \mu _{\alpha \dot
\alpha }\thetabar^ \alphadot \partial_\mu\Bigr) \phi_{\rm ac} (x, \theta,
\thetabar) \ , \\
\bar \delta_\alphadot \phi_{\rm ac}(x, \theta, \thetabar) &  =  &  -
\frac{\partial}{ \partial\theta^\alphadot}  \phi_{\rm ac} (x, \theta,
\thetabar) \ .\nonumber 
\end{eqnarray}
\end{itemize}

\paragraph{Superspace integration}

\begin{eqnarray}
\intV  \phi & = & \intd \frac{\partial}{\partial\theta_\alpha}
\frac{\partial}{\partial\theta^\alpha}
\frac{\partial}{\partial\thetabar^ \alphadot}
\frac{\partial}{\partial\thetabar _\alphadot} \phi  \ ,\\
\intS  A & = & \intd \frac{\partial}{\partial\theta_\alpha}
\frac{\partial}{\partial\theta^\alpha}
A  \ , \qquad 
\intSbar \bar  A  =  \intd \frac{\partial}{\partial\theta^ \alphadot}
\frac{\partial}{\partial\theta_\alphadot}
 \bar A \ ,\nonumber
\end{eqnarray}
where $\phi $  is a real superfield and $A$ and $\bar A $ are chiral
and antichiral superfields respectively.

\section{Discrete symmetries}

SQED with local coupling and gauged axial symmetry is invariant under
the discrete symmetries parity P, charge conjugation C and R-parity.
We define the transformation of the fields according to the following
table:
\begin{table}[tbh]
\begin{displaymath}
\begin{array}{|c||c|c|c|c|c|c|c|c|c|c|c|c|c|c|c|c|c|c|}
\hline
 & x^\mu & A^\mu  & \lambda^\alpha &  
\varphi_{L} & \varphi_{R} &
\psi^\alpha_{L} &  
\psi^\alpha_{R} &   c & \epsilon^\alpha & 
\omega^\nu & \bar{c} & B 
\\ \hline
{\rm R} & x^\mu & A^\mu & - i\lambda^\alpha & 
-i\varphi_L & -i\varphi_R & 
\psi^\alpha_{L} & 
\psi^\alpha_{R} &   c & -i\epsilon^\alpha & 
\omega^\nu & \bar{c} & B 
\\ \hline
{\rm C} & x^\mu & -A^\mu  & - \lambda^\alpha &   
\varphi_{R} & \varphi_{L} &
\psi^\alpha_{R} &  
\psi^\alpha_{L} &  -c & \epsilon^\alpha & 
\omega^\nu & -\bar{c} & -B 
\\ \hline
{\rm CP} & ({\cal P}x)^\mu & -({\cal P}A)^\mu  & - i \lambdabar_\alphadot &
\phibar_{L} &\phibar_{R} &
i\psibar_L{}_\alphadot &  
i\psibar_R{}_\alphadot &   -c & -i\epsilonbar_\alphadot &
 ({\cal P}\omega)^\nu & -\bar{c} & -B 
\\ \hline
\end{array}
\end{displaymath}
\end{table}

\begin{table}[tbh]
\begin{displaymath}
\begin{array}{|c||c|c|c|c|c|c|c|c|c|c|c|c|c|c|c|c|c|c|}
\hline
   &  ~e~ &~\eta- \etabar~& ~\chi^ \alpha~ &~f~ &
~q~& q ^\alpha & ~q_F~ & V^\mu  & \tilde \lambda^\alpha & ~\tilde D~ & ~\tilde c~
\\ \hline
{\rm R} &  e&  \eta -\etabar & i \chi^ \alpha& -f & q & i q ^\alpha & -q_F &
   V^\mu & - i \tilde \lambda^\alpha & \tilde D & \tilde c 
\\ \hline
{\rm C}  & e & \eta - \etabar & \chi^ \alpha & f & q  & q^ \alpha & q_F &
V^\mu  & \tilde \lambda^\alpha & \tilde D& \tilde c   
\\ \hline
{\rm CP} & e & \etabar - \eta & i \chibar_\alphadot & \bar  f & 
\qbar & i \qbar_\alphadot & \qbar _F & -({\cal P}V)^\mu  & - i \lambdaVbar_\alphadot &
\tilde D &  - \tilde c  
\\ \hline
\end{array}
\end{displaymath}
{\caption{{Discrete symmetries.}
The transformation rules for the sources $Y_i$ can be deduced from the
requirement that $\Gamma_{\rm ext}$ is invariant. The
transformation rules for the complex conjugate fields are obvious
except for the $CP$ conjugation of the spinors. We define 
for $\chi\in\{\lambda,\psi_L,\psi_R,\epsilon\}:$
\hspace*{\fill}
\mbox{\hspace*{0cm}}
\hspace*{\fill}
\mbox{$\chi^\alpha \stackrel{CP}{\to} a \chibar_\alphadot\Rightarrow
\chibar_\alphadot \stackrel{CP}{\to} -a^* \chi^\alpha \ ,\ 
\chi_\alpha \stackrel{CP}{\to} -a \chibar^\alphadot\ , \ 
\chibar^\alphadot \stackrel{CP}{\to} a^* \chi_\alpha 
\ .$}
\hspace*{\fill}
\label{TabDiscreteSym}
}}
\end{table}

\section{The BRS transformations} 
 
In this appendix we list the BRS transformations of all fields
introduced in SQED with local gauge coupling and with 
the axial vector multiplet. 
\begin{itemize}
\item BRS transformations of the photon multiplet
\begin{eqnarray}
sA_\mu & = & \frac 1e \partial_\mu e c + i\epsilon\sigma_\mu\lambdabar
             -i \lambda\sigma_\mu\epsilonbar \\
 &  &  +
\frac 12 e ^2(\epsilon \chi + \chibar
\epsilonbar)A_\mu-i\omega^\nu\partial_\nu A_\mu 
\ ,\nonumber \\
s\lambda^\alpha & = & \frac{i}{2e} (\epsilon\sigma^{\rho\sigma})^\alpha
             F_{\rho\sigma}(eA) - i\epsilon^\alpha\,
             eQ_L(|\phi_L|^2-|\phi_R|^2) + \frac 12 \epsilon^ \alpha
e^2 ( \chi \lambda - \chibar \lambdabar)\nonumber \\
& & + \frac 12 e ^2(\epsilon \chi + \chibar \epsilonbar) \lambda^ \alpha
              -i\omega^\nu\partial_\nu  
             \lambda^\alpha
\ ,\nonumber \\
s\lambdabar_\alphadot & = & \frac{-i}{2e} (\epsilonbar\sigmabar^{\rho\sigma})
             _\alphadot F_{\rho\sigma}(eA) - i\epsilonbar_\alphadot\, 
             eQ_L(|\phi_L|^2-|\phi_R|^2) 
+ \frac 12 \epsilonbar_ \alphadot
e^2 ( \chi \lambda - \chibar \lambdabar)\nonumber \\
& & + \frac 12 e ^2(\epsilon \chi + \chibar \epsilonbar) \lambdabar_ \alphadot
             -i\omega^\nu\partial_\nu \lambdabar_\alphadot \ . \nonumber
\end{eqnarray}
\item BRS transformations of the axial vector multiplet
\begin{eqnarray}
sV_\mu & = & \partial_\mu \tilde c + i\epsilon\sigma_\mu{\bar {\tilde
\lambda}}
             -i \tilde
\lambda\sigma_\mu\epsilonbar -i\omega^\nu\partial_\nu V_\mu
\ ,\\
s\tilde\lambda^\alpha & = & \frac{i}{2} (\epsilon\sigma^{\rho\sigma})^\alpha
             F_{\rho\sigma}( V) + \frac i2 \epsilon^\alpha\,
             \tilde D -i\omega^\nu\partial_\nu  
             \tilde \lambda^\alpha
\ ,\nonumber \\
s{\bar {\tilde \lambda}}
_\alphadot & = & \frac{-i}{2} (\epsilonbar\sigmabar^{\rho\sigma})
             _\alphadot F_{\rho\sigma}(V) + \frac i 2 \epsilonbar_\alphadot 
\tilde D
             -i\omega^\nu\partial_\nu \bar {\tilde \lambda}_\alphadot
\ , \nonumber \\
s \tilde D & = & 2 \epsilon \sigma^ \mu \partial_\mu {\overline {\tilde \lambda}} + 
                  2 \partial_
 \mu  {\tilde \lambda} \sigma^ \mu \epsilonbar 
             -i\omega^\nu\partial_\nu  {\tilde D} \ .\nonumber
\end{eqnarray}
\item BRS transformations of  matter fields \\
 The covariant                   derivative is
defined by eq. (\ref{covDV}).
\begin{eqnarray}
s\varphi_L & = & -i(eQ_L c + \tilde c)\varphi_L +\sqrt2\,
\epsilon\psi_L
 - i\omega^\nu\partial_\nu \varphi_L
\ ,\nonumber \\
s\phibar_L & = & +i(eQ_L c+ \tilde c)\,\phibar_L   +
\sqrt2\, \psibar_L\epsilonbar - i\omega^\nu\partial_\nu \phibar_L
\ ,\nonumber \\
s\psi_L^\alpha & = & -i(eQ_L c+ \tilde c)\,\psi_L^\alpha
\nonumber \\
 & &  - \sqrt2\, 
         \epsilon^\alpha\, (\qbar+m)\phibar_R
         -\sqrt2\, i (\epsilonbar\sigmabar^\mu)^\alpha D_\mu\varphi_L 
         -i\omega^\nu\partial_\nu \psi_L^\alpha
\ ,\nonumber \\
s{\psibar_L}_\alphadot & = & +i(eQ_L c + \tilde c)\,{\psibar_L}_\alphadot 
\nonumber \\
& &          + \sqrt2\,\epsilonbar_\alphadot\,(q+ m)\varphi_R + \sqrt2\, i
         (\epsilon\sigma^\mu)_\alphadot D_\mu\phibar_L
         -i\omega^\nu\partial_\nu {\psibar_L}_\alphadot \ .\nonumber
\end{eqnarray}
and respective expressions for right-handed fields.
\item The BRS transformations of ghosts
\begin{eqnarray}
sc & = & 2i\epsilon\sigma^\nu\epsilonbar A_\nu +
\frac 12 e^2 (\epsilon \chi + \chibar \epsilonbar) c
-i\omega^\nu\partial_\nu c
\ ,\\
s\tilde  c & = & 2i\epsilon\sigma^\nu\epsilonbar V_\nu
-i\omega^\nu\partial_\nu\tilde  c
\ ,\nonumber \\
s\epsilon^\alpha & = & 0
\ ,\nonumber \\
s\epsilonbar^\alphadot & = &0
\ ,\nonumber \\
s\omega^\nu & = & 2\epsilon\sigma^\nu\epsilonbar \ . \nonumber
\end{eqnarray}
\item BRS transformations of the $B$-field, anti-ghost and gauge parameter\\
\begin{eqnarray}
\label{BRSgaugefixing}
s B &  = &
2i \epsilon \sigma ^ \nu \epsilonbar \partial_\nu (\frac 1e
\cbar) - e^2 \frac 12  (\epsilon \chi + \chibar \epsilonbar )B - i
\omega^ \nu
\partial_ \nu B \ ,\\
s \cbar &  = & B
 - e^2 \frac 12  (\epsilon \chi + \chibar \epsilonbar ) \cbar - i
\omega^ \nu
\partial_ \nu \cbar \ , \nonumber \\
s \chi_\xi &  = &
2i\frac 1{e^ 2} \epsilon \sigma ^ \nu \epsilonbar \partial_\nu( e ^2 
\xi)  + e^2   (\epsilon^ \alpha \chi_\alpha + \chibar_\alphadot
 \epsilonbar^ \alphadot ) \chi_\xi - i
\omega^ \nu
\partial_ \nu \chi_\xi \ ,\nonumber \\
s \xi &  = & \chi_\xi
 +e^2   (\epsilon \chi + \chibar \epsilonbar ) \xi - i
\omega^ \nu
\partial_ \nu \xi \ .\nonumber
\end{eqnarray}
\item BRS transformations of the local coupling and its superpartners (\ref{E2def})
\begin{eqnarray}
s \eta & = & \epsilon^ \alpha \chi_\alpha -i \omega ^ \nu \partial_\nu \eta \ ,\\
s \etabar & = & \chibar_\alphadot \epsilonbar^ \alphadot
 -i \omega ^ \nu \partial_\nu
\etabar \ ,\nonumber \\
s \chi_\alpha& = & 2 i (\sigma^ \mu \epsilonbar)_\alpha \partial_\mu \eta + 2
\epsilon_\alpha f  
- i \omega^ \mu \partial_\mu
\chi_\alpha \ ,\nonumber \\
s \chibar_\alphadot& = & 2 i (\epsilon \sigma^ \mu)_\alphadot
 \partial_\mu \etabar   - 2
\epsilonbar_\alphadot \bar f 
- i \omega^ \mu \partial_\mu
\chibar_\alphadot \ ,\nonumber \\
s f & = &   i 
\partial_\mu \chi \sigma^ \mu \epsilonbar - i \omega^ \mu \partial_\mu
f \ ,\nonumber \\
s \bar f & = & - i 
 \epsilon \sigma^ \mu \partial_\mu \chibar 
- i \omega^ \mu \partial_\mu
\bar f \ . \nonumber
\end{eqnarray}
\item BRS transformations of $q$-multiplets (\ref{qdef})\\
 The covariant derivative is defined in eq.~(\ref{covq})
\begin{eqnarray}
s q & = &  + 2i \tilde c (q  + m ) +
 \epsilon^ \alpha q_\alpha -i \omega ^ \nu \partial_\nu q\ ,\\
s \qbar & = & -2i \tilde c (\qbar + m)
+ \qbar_\alphadot \epsilonbar^ \alphadot
 -i \omega ^ \nu \partial_\nu
\qbar \ , \nonumber \\
s q_\alpha& = & + 2i \tilde c q_\alpha +2 i (\sigma^ \mu \epsilonbar)_\alpha 
D_\mu (q +m) + 2
\epsilon_\alpha q_F 
- i \omega^ \mu \partial_\mu
q_\alpha \ ,\nonumber \\
s \qbar_\alphadot& = &  - 2i \tilde c \qbar_\alphadot +
2 i (\epsilon \sigma^ \mu)_\alphadot
 D_\mu (\qbar +m )   - 2
\epsilonbar_\alphadot \qbar_F
- i \omega^ \mu \partial_\mu
\qbar_\alphadot \ ,\nonumber \\
s q_F & = &+ 2i \tilde c q_F +
 i 
D_\mu q^ \alpha \sigma^ \mu_{\alpha \alphadot} \epsilonbar^ \alphadot
- 4 i
 \lambdaVbar _ \alphadot \epsilonbar^ \alphadot (q +m )
 - i \omega^ \mu \partial_\mu
q_F \ , \nonumber \\
s \qbar_F & = & - 2i \tilde c \qbar_F  - i 
 \epsilon^ \alpha \sigma^ \mu_{\alpha \alphadot} D_\mu \qbar^ \alphadot 
+ 4 i
\epsilon ^ \alpha \lambdaV _ \alpha  (\qbar + m)
- i \omega^ \mu \partial_\mu
\qbar_ F \ . \nonumber
\end{eqnarray}
\end{itemize}

\end{appendix}

\end{document}